\documentclass[prd,superscriptaddress,a4paper,showpacs,showkeys,10pt,nofootinbib]{revtex4}
\usepackage{graphicx}
\usepackage{graphics}
\usepackage{epsfig}
\usepackage{dcolumn}
\usepackage{bm}
\usepackage{multirow}
\usepackage{tabularx}
\usepackage{hyperref}
\usepackage{commath}
\usepackage{epstopdf}
\usepackage[T1]{fontenc}
\usepackage{geometry}
\usepackage{color}

\providecommand{\superchic}{{\sc superchic}}

\newcommand{\sqrtsnn}{\sqrt{s_{_{\textsc{nn}}}}}

\def\be{\begin{equation}}
\def\ee{\end{equation}}

\providecommand{\ee}{e$^+$e$^-$}

\providecommand{\tabularnewline}{\\}

\makeatother

\begin{document}
%
%
\title{Production of axionlike particles in $PbPb$ collisions at the LHC, HE -- LHC and FCC: A phenomenological analysis}


\author{R. O. Coelho}

\email[]{coelho72@gmail.com}

\affiliation{Instituto de F\'{\i}sica e Matem\'atica, Universidade Federal de
Pelotas (UFPel),\\
Caixa Postal 354, CEP 96010-090, Pelotas, RS, Brazil}

\author{V. P. Gon\c calves}

\email[]{barros@ufpel.edu.br}

\affiliation{Instituto de F\'{\i}sica e Matem\'atica, Universidade Federal de
Pelotas (UFPel),\\
Caixa Postal 354, CEP 96010-090, Pelotas, RS, Brazil}

\author{D. E. Martins}

\email[]{dan.ernani@gmail.com}

\affiliation{Instituto de F\'isica, Universidade Federal do Rio de Janeiro (UFRJ), 
Caixa Postal 68528, CEP 21941-972, Rio de Janeiro, RJ, Brazil}

\author{M. S. Rangel}

\email[]{rangel@if.ufrj.br}

\affiliation{Instituto de F\'isica, Universidade Federal do Rio de Janeiro (UFRJ), 
Caixa Postal 68528, CEP 21941-972, Rio de Janeiro, RJ, Brazil}



\begin{abstract}
The production of axionlike particles (ALP) in ultraperipheral $PbPb$ collisions (UPHIC)  is investigated considering the energies of the next run of the Large Hadron Collider (LHC) and of the future High  Energy -- LHC (HE -- LHC) and Future Circular Collider (FCC).  Assuming four different combinations for the ALP mass and coupling and the typical exclusivity cuts for central and forward detectors, we estimate the cross section and invariant mass, rapidity, transverse momentum and {acoplanarity} distributions associated to the diphoton final state produced in the $\gamma \gamma \rightarrow a \rightarrow \gamma \gamma$ subprocesses. A detailed analysis of the backgrounds is performed. We demonstrate that  the backgrounds can be strongly reduced by the exclusivity cuts and that a forward detector, as the LHCb, is ideal to probe an ALP with small mass. Finally, our results indicate that a future experimental analysis of the diphoton final state in UPHIC can probe the existence and properties of  axionlike particles.
\end{abstract}




\maketitle

\section{Introduction}
\begin{figure}
\begin{tabular}{cc}
\hspace{-1cm}
{\psfig{figure=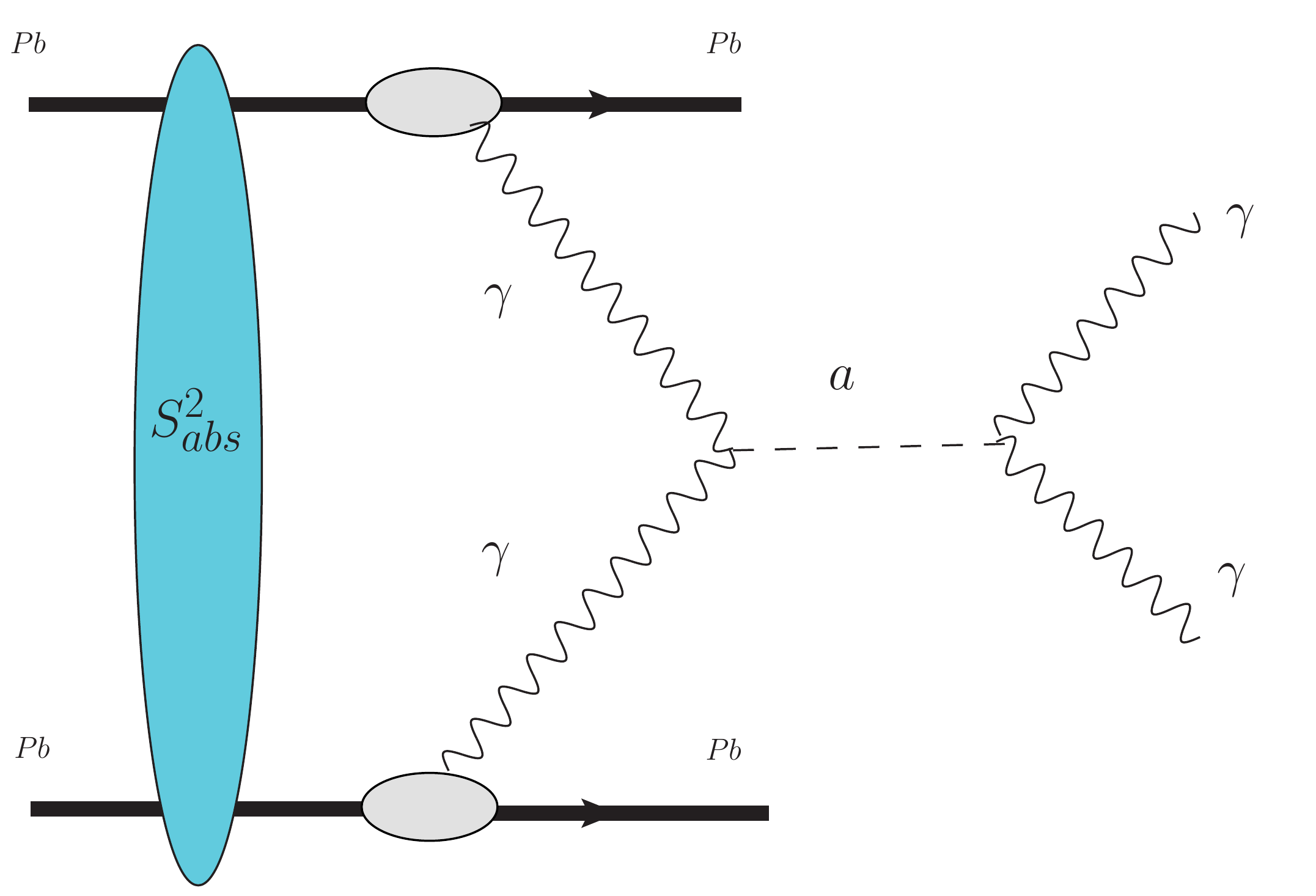,width=8.0cm}} &
{\psfig{figure=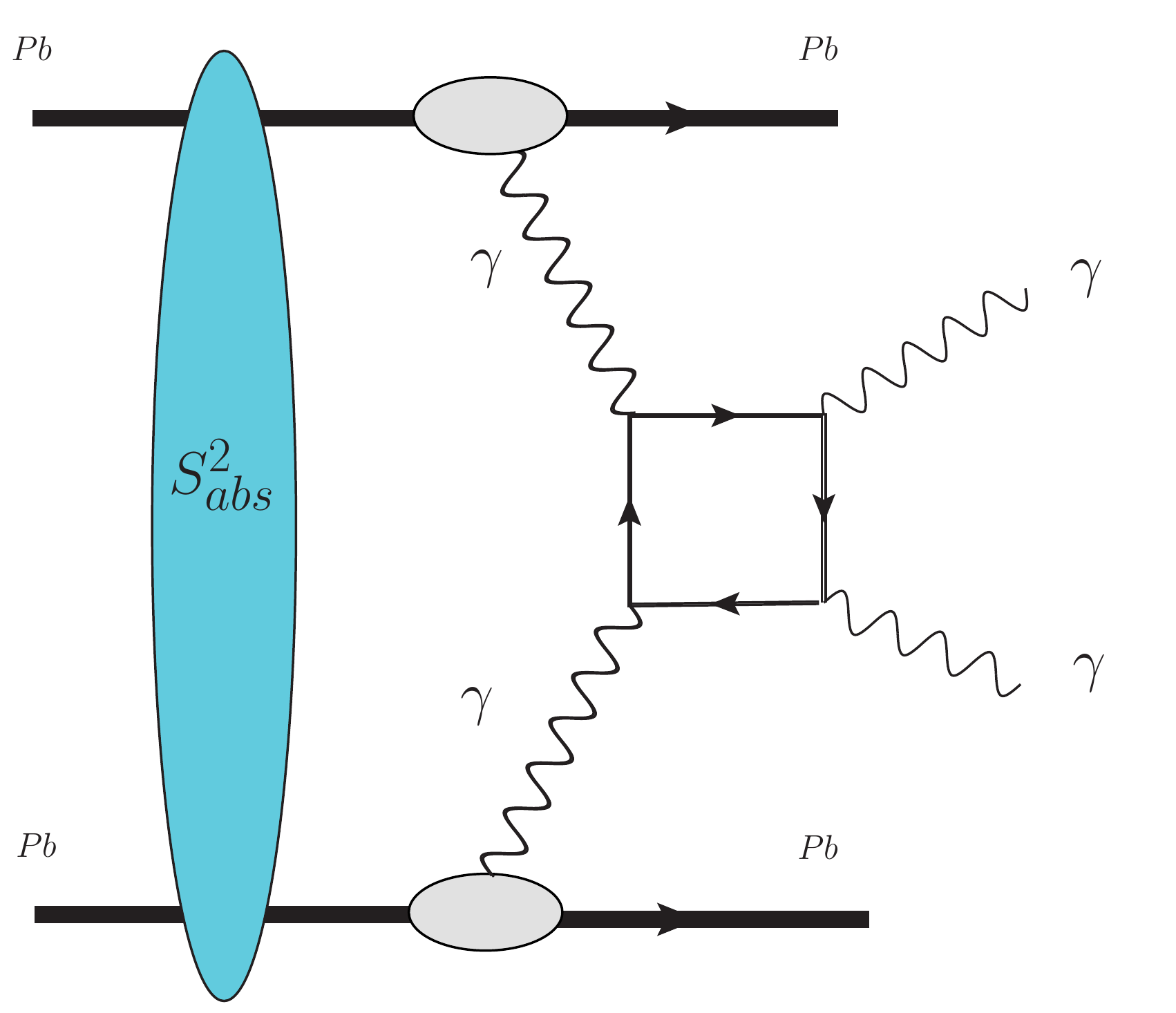,width=6.2cm}}  \\
(a) & (b) \\ 
\end{tabular}                                                                                                                       
\caption{Diphoton production in $PbPb$ collisions by (a) the $\gamma \gamma \rightarrow a \rightarrow\gamma \gamma$ subprocess and (b) Light -- by -- Light scattering.}
\label{fig:diagram}
\end{figure}

Over  the last few years there has been a rising interest in searching for axionlike particles in $e^+e^-$, $ep$, $\nu p$, $pp$, $pA$ and $AA$ collisions as well in laser beam experiments 
(See e.g. Refs. \cite{Jaeckel:2015jla,Bauer:2017ris,knapen,Aloni:2018vki,royon,Aloni:2019ruo,Bauer:2018uxu,Yue:2019gbh,Ebadi:2019gij,Alves:2019xpc}), mainly motivated by the fact that such particles are  predicted to occur in many extensions of the Standard Model (SM).
They are pseudo -- Nambu -- Goldstone bosons, which arise in models with spontaneous breaking of a global symmetry and are expected to be characterized by a small mass in comparison to the scale of the spontaneous breaking  and by couplings to the Standard Model (SM) particles that are, at least, suppressed by the inverse of the same scale. Depending on the ALP mass and coupling structure, they can be produced at colliders and decay into photons, charged leptons, light hadrons or jets, which can be detected. In our analysis we are particularly interested in the coupling of  the pseudoscalar ALP $a$ to photons, which is described by a Lagrangian of the form
\begin{equation}
\mathcal{L}=\frac{1}{2}\partial^\mu a \partial_\mu a -\frac{1}{2}m_a^2 a^2 -\frac{1}{4}g_a a F^{\mu\nu}\tilde{F}_{\mu\nu}\;,
\end{equation}
where $m_a$ is the ALP mass, $g_a$ is the coupling constant and  $\tilde{F}^{\mu\nu}  = \frac{1}{2}\epsilon^{\mu\nu \alpha\beta} F_{\alpha\beta}$. As a consequence, the ALP can be produced by the photon -- photon fusion and can decay into a diphoton system. In Ref. \cite{knapen} the authors have proposed to search by axionlike particles in ultraperipheral heavy ion collisions (UPHIC)~[See also Ref. \cite{royon}], which  are characterized by an impact parameter $b$ greater than the sum of the radius of the colliding  nuclei \cite{upc1,upc2,upc3,upc4,upc5,upc6,upc7,upc8,upc9} and by a photon -- photon luminosity that scales with $Z^4$, where $Z$ is number of protons in the nucleus. The ALP production in UPHIC
 is represented in Fig. \ref{fig:diagram} (a) and the associated cross section can be derived using the equivalent photon approximation \cite{epa}. In this approach, we can associated to the incident nucleus an equivalent photon spectrum 
 $N(\omega_i, {\mathbf r}_i)$, which allows to estimate the number the photons  with energy $\omega_i$ at a transverse distance ${\mathbf r}_i$  from the center of nucleus, defined in the plane transverse to the trajectory, which is determined by the charge form factor of the nucleus. Consequently, 
 the total cross section can be factorized in terms of the equivalent photon spectrum  
of the incident nuclei and the elementary cross section for the $\gamma \gamma \rightarrow a \rightarrow\gamma \gamma$ process as follows
\begin{eqnarray}
\sigma \left(Pb Pb \rightarrow Pb \otimes \gamma \gamma \otimes Pb;s \right)   
&=& \int \mbox{d}^{2} {\mathbf r_{1}}
\mbox{d}^{2} {\mathbf r_{2}} 
\mbox{d}W 
\mbox{d}y \frac{W}{2} \, \hat{\sigma}\left(\gamma \gamma \rightarrow a \rightarrow \gamma \gamma ; 
W \right )  N\left(\omega_{1},{\mathbf r_{1}}  \right )
 N\left(\omega_{2},{\mathbf r_{2}}  \right ) S^2_{abs}({\mathbf b})  
  \,\,,
\label{cross-sec-2}
\end{eqnarray}
where $\sqrt{s}$ is center - of - mass energy of the $PbPb$ collision, $\otimes$ characterizes a rapidity gap in the final state, 
$W = \sqrt{4 \omega_1 \omega_2} = m_X$ is the invariant mass of the $\gamma \gamma$ system  and $y$ its rapidity. Moreover, in order to exclude the overlap between the colliding nuclei and insure the dominance of the electromagnetic interaction, it is useful to include in Eq.(\ref{cross-sec-2}) the absorptive factor $S^2_{abs}({\mathbf b})$, which depends on the impact parameter ${\mathbf b}$ of the $PbPb$ collision.    One of the main advantages of the ALP search in UPHIC is that the resulting  final state is very clean, consisting  of the diphoton system,  two intact nuclei and  two rapidity gaps, i.e. empty regions  in pseudo-rapidity that separate the intact very forward nuclei from the $\gamma \gamma$ system. However, in order to probe the ALP in the $\gamma \gamma \rightarrow a \rightarrow\gamma \gamma$ channel, it is fundamental to disentangle the associated events  from those generated in the Light -- by -- Light (LbL) scattering, in which the diphoton final state is created by the elementary  elastic  $\gamma \gamma \rightarrow \gamma \gamma$ subprocess, represented in Fig. \ref{fig:diagram} (b). As demonstrated in Ref. \cite{nosdiphoton}, where the diphoton production by the LbL, Durham and double diffractive processes was estimated, the LbL process dominates the diphoton production at small invariant masses when the exclusivity cuts (see below) are taking into account.
Our goal in this paper is twofold. First, to present, for the first time, a detailed analysis of the ALP production in the kinematical range probed by the LHCb detector, which is expected to be able to probe ALP's with smaller invariant masses than the central detectors. Second, 
to present predictions for the ALP production in $PbPb$ collisions for the energies of the   High -- Energy LHC ($\sqrt{s} = 10.6$ TeV) \cite{he_lhc} and Future Circular Collider ($\sqrt{s} = 39$ TeV) \cite{fcc}  considering  the typical configurations of  central and forward detectors and similar cuts to those used to LHC. In our analysis of the signal and the LbL background we will use \superchic 3 Monte Carlo event generator \cite{superchic3}, which has been recently generalized to treat ion -- ion collisions.


This paper is organized as follows. In the next Section,  we present our results for the ALP production at the LHC, HE -- LHC and FCC. Predictions for cross sections and  the invariant mass, rapidity, transverse momentum  and acoplanarity distributions are presented. The impact of the selection cuts is discussed and predictions for  typical central and forward detectors are presented.  Finally, in Section \ref{sec:conc}, our main conclusions are summarized.

\section{Results}
\label{sec:res}

Following Ref.\cite{superchic3}, we will assume that the photon spectrum can be expressed in terms of the electric form factor  and that the absorptive corrections $S^2_{abs}({\mathbf b})$ for $\gamma \gamma$ interactions  can be estimated taking into account the multiple scatterings between the nucleons of the incident nuclei, which allow to calculate the probability for no additional  ion -- ion rescattering at different impact parameters. For the background associated to the LbL scattering, the  elementary cross section 
$\hat{\sigma}(\gamma \gamma \rightarrow \gamma \gamma)$ will be calculated taking into account of the fermion loop contributions as well as  the contribution from $W$ bosons. In addition, we also will present the predictions for the backgrounds associated to the diphoton production by the Durham and double diffractive processes (DDP), obtained originally in Ref. \cite{nosdiphoton}, which are here complemented by the inclusion of new cuts on the invariant mass of the diphoton system. For a detailed discussion about the Durham and DDP channels we refer the reader to the Ref. \cite{nosdiphoton}. On the other hand, the signal associated to the ALP production will be calculated following the approach discussed in Ref. \cite{superchic3}, where the $\gamma \gamma \rightarrow a \rightarrow \gamma \gamma$ cross sections is estimated assuming that the ALP is a narrow resonance with a mass $m_a$ that couples to the $\gamma \gamma$ system with strength $g_a$.

\begin{center}
\begin{table}[t!]
\begin{tabularx}{\textwidth}{@{}l *4{>{\centering\arraybackslash}X}@{}}
\hline\hline
\multirow{2}{*}{Subprocess} & \multirow{2}{*}{$\sqrt{s}$ (TeV)} & \multicolumn{3}{c}{$\sigma[Pb\:Pb\to  Pb+ \gamma \gamma +Pb ]$} \\
          &               &  ALP mass (GeV)  & Coupling (GeV$^{-1}$)         & \superchic3      \\
            \hline
\multirow{2}{*}{$\gamma \gamma \to a \to  \gamma \gamma$}
       & 5.5            &   3     & $1.0\times10^{-3}~$      &  $1.3\times10^{4}$~nb   \\ 
    & 10.6            &        &       &  $2.1\times10^{4}$~nb   \\ 
        & 39             &         &    &  $4.3\times10^{4}$~nb     \\
\hline
        & 5.5            &   5     & $2.0\times10^{-4}$      & 363.0~nb  \\
          & 10.6            &        &       &  587.4~nb   \\ 
        & 39             &         &     &   $1300.0$~nb  \\
\hline
        & 5.5            &   15    & $0.06\times10^{-3}$      & $11.0$~nb  \\
          & 10.6            &        &       &  21.7~nb   \\ 
        & 39             &         &      & $61.0$~nb    \\
\hline
        & 5.5            &   40    & $1.3\times10^{-4}$      & $13.0$~nb  \\ 
          & 10.6            &        &       &  35.1~nb   \\ 
        & 39             &         &    & $140.0$~nb    \\
        \hline
  \hline
\end{tabularx}
\caption{Predictions for the ALP cross sections considering $PbPb$ collisions at $\sqrt{s} = 5.5,\, 10.6$ and 39 TeV and four different combinations for the values of the ALP mass $m_a$ and the coupling $g_a$.}
\label{tab:generation} 
\end{table}
\end{center}

\begin{center}
 \begin{figure}[t!]
 \includegraphics[width=0.45\textwidth]{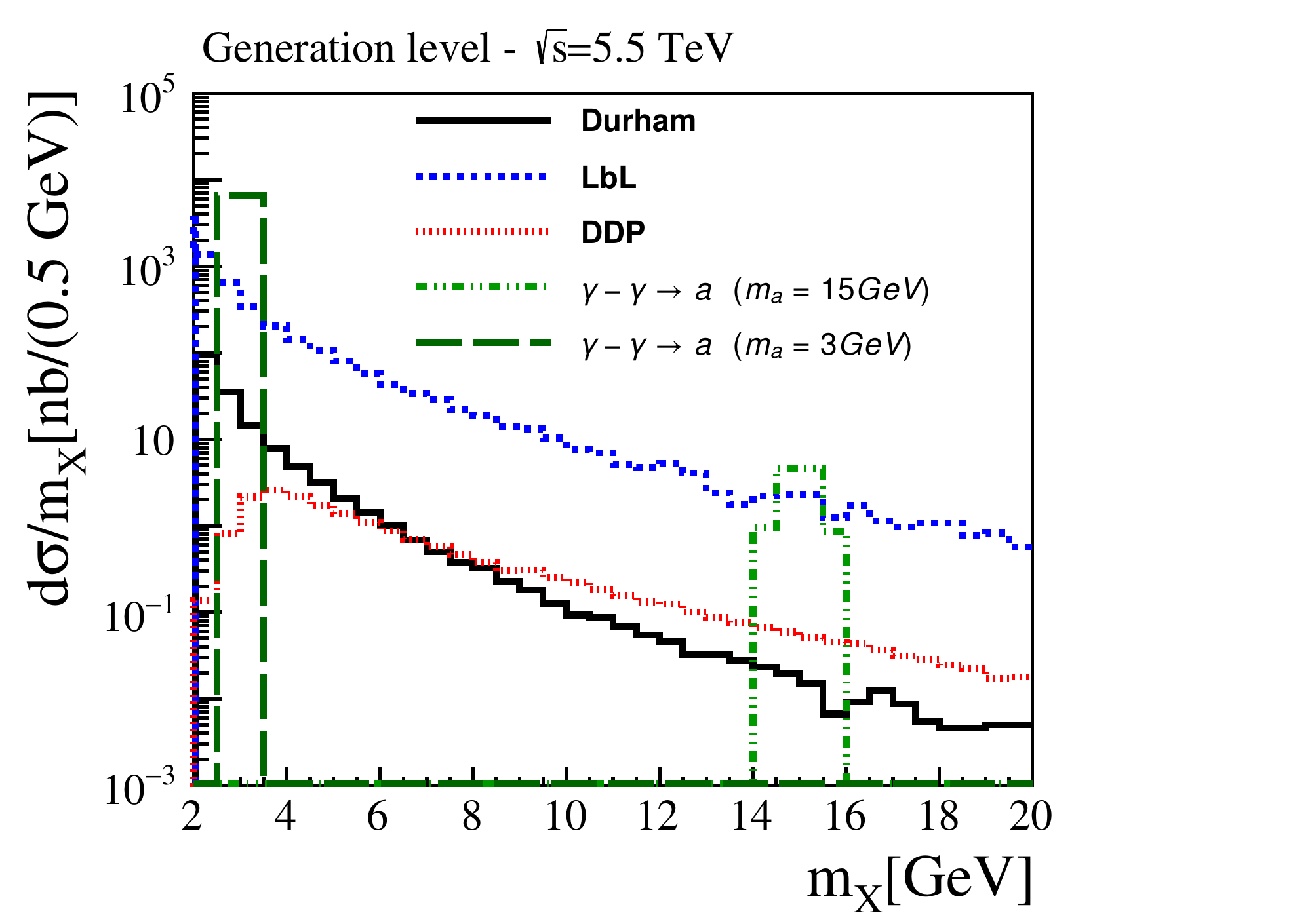}
  \includegraphics[width=0.45\textwidth]{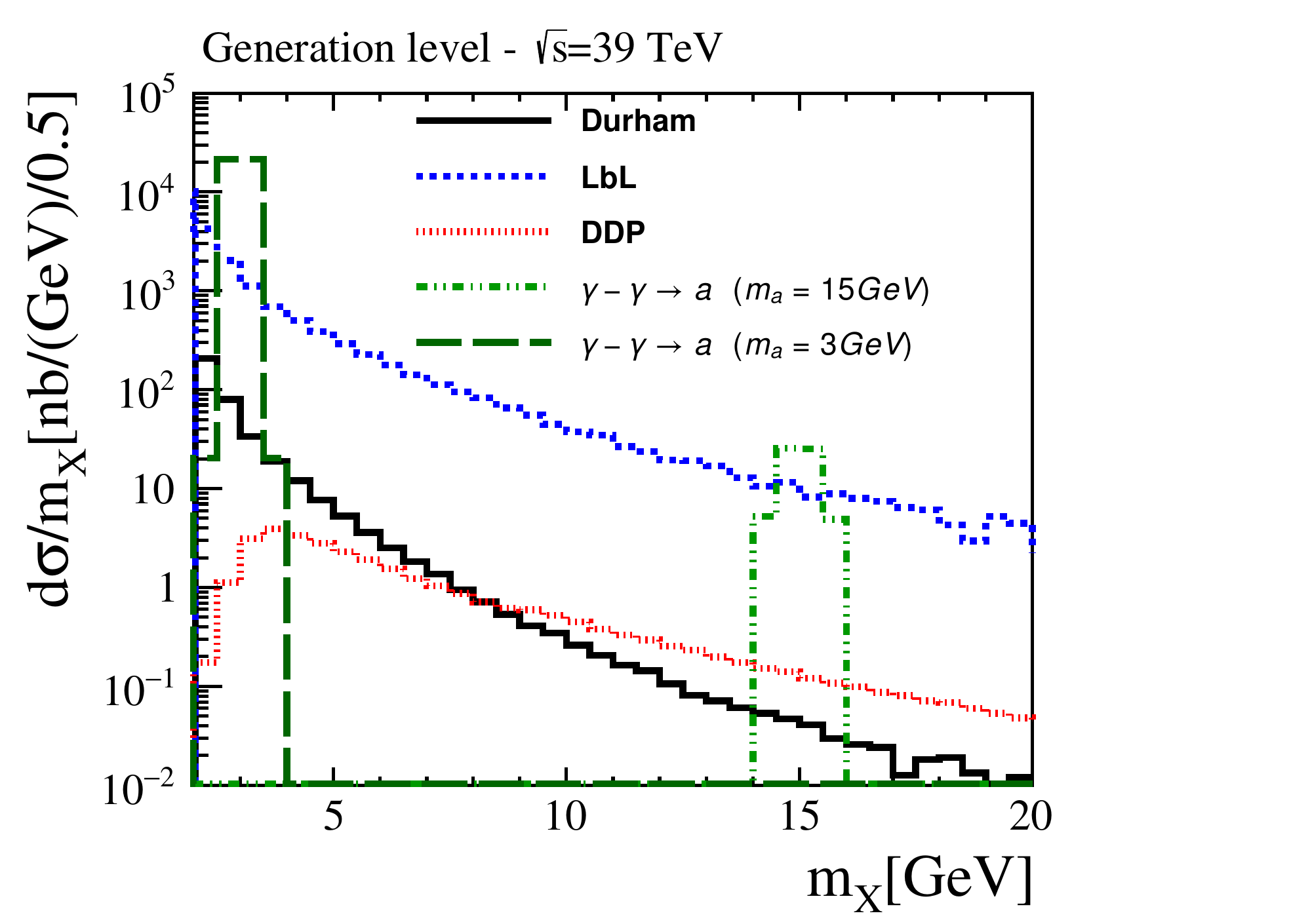}
 \includegraphics[width=0.45\textwidth]{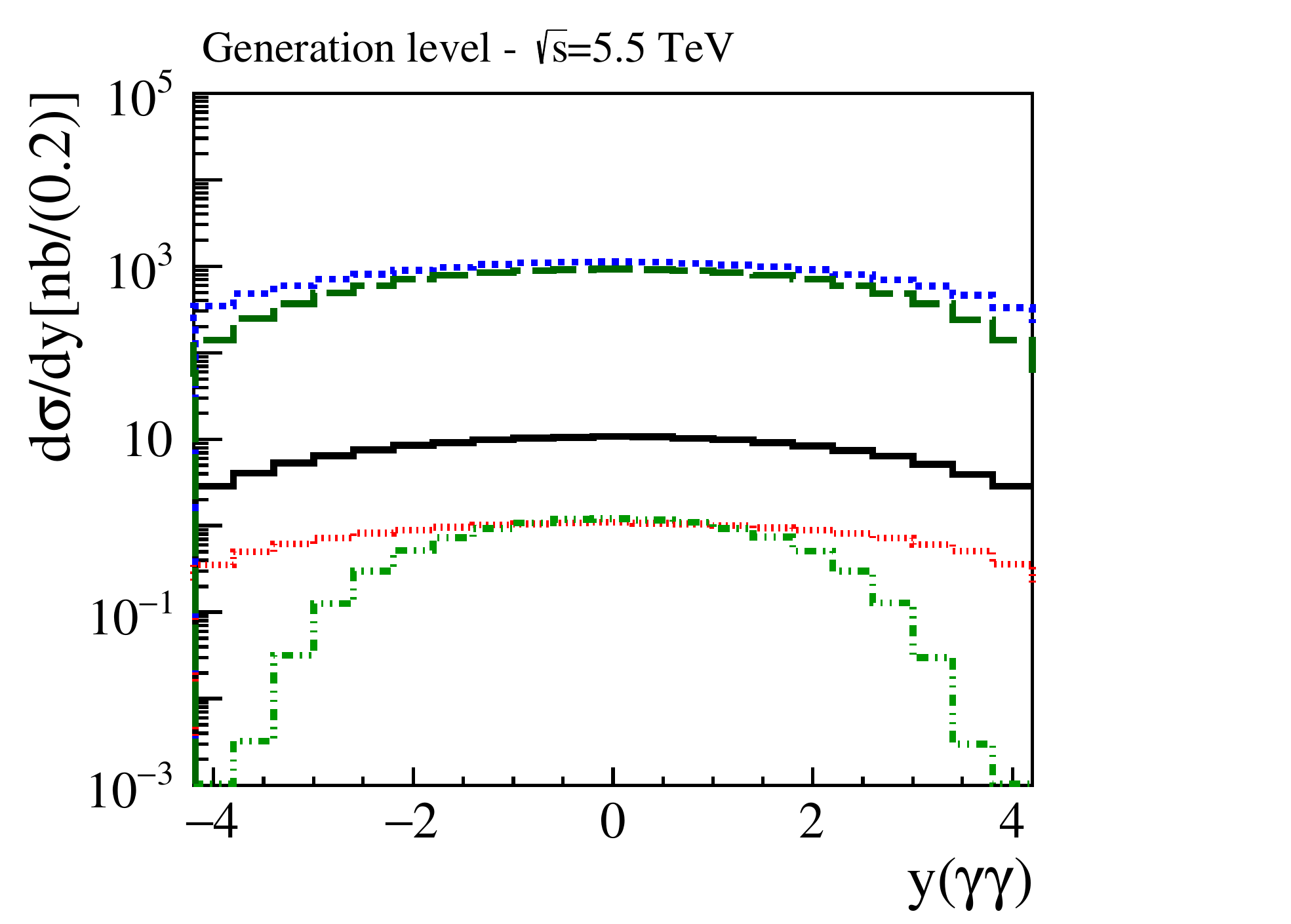}
 \includegraphics[width=0.45\textwidth]{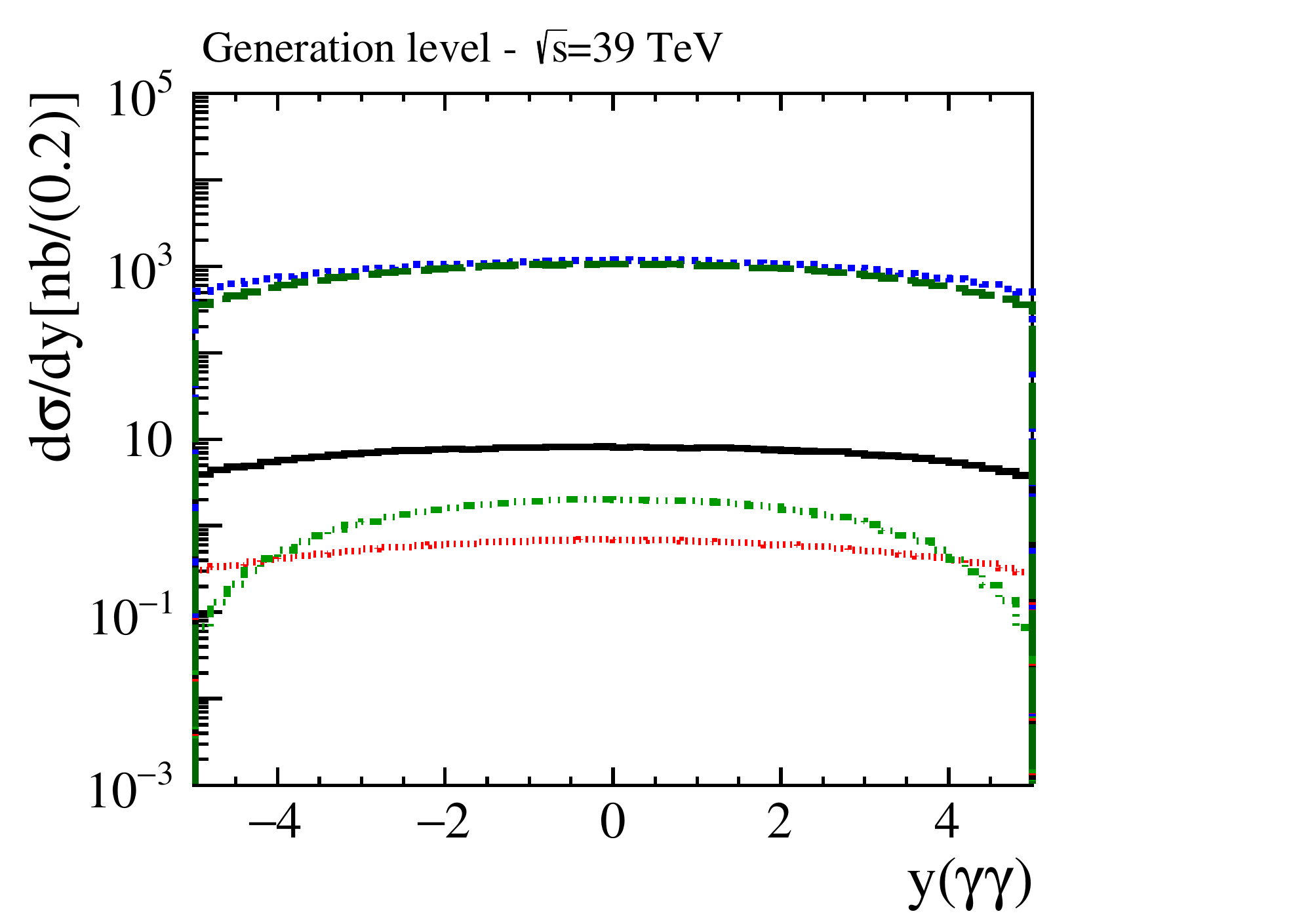}
\caption{Differential cross sections as function of invariant mass $m_{X}$ and rapidity $y({\gamma\gamma})$ of the diphoton system in $PbPb$ collisions for LHC (left panels) and FCC (right panels) energies. Results at generation level, without the inclusion of experimental cuts.}
\label{fig:generation}
 \end{figure}
 \end{center}

In what follows we will present our results for the ALP production in $PbPb$ collisions at $\sqrt{s} = 5.5$, 10.6 and 39 TeV. In our analysis we will use the \superchic 3 MC event generator \cite{superchic3} to estimate the processes represented in the Figs. \ref{fig:diagram} (a) and (b). 
We will consider the following representative combinations of axion mass and coupling: $(m_a; \, g_a) = (3.0; \, 1.0 \times 10^{-3}), \, (5.0; \, 2.0 \times 10^{-4}), \, (15.0; \, 0.06 \times 10^{-4})$ and $(40.0; \, 1.3 \times 10^{-4})$ in units of (GeV; GeV$^{-1}$). 
Initially, in Table \ref{tab:generation} we  present our results for the  ALP cross sections obtained at the generation level, without the inclusion of any selection in the events. 
We have that the cross section increases for smaller masses and larger energies, being of the order of $\mu b$ at $m_X = 3.0$ GeV and FCC energy. For comparison, we have that the LbL cross sections at  
$\sqrt{s} = 5.5/\, 10.6/\, 39$ TeV are $1.8/\,2.7/\,5.2 \times 10^4$ nb, respectively. Therefore, our results indicate that the ALP cross section can be of the same order of the LbL one at small -- $m_a$ and is non -- negligible for larger masses. In Fig. \ref{fig:generation} we present our predictions for the invariant mass and rapidity distributions of the diphoton system, derived at the generation level considering two possible axion masses and $PbPb$ collisions at the LHC (left panels) and FCC (right panel) energies. The predictions for the diphoton production by the Durham and double diffractive processes, obtained taking into account of the soft survival corrections as derived in Ref. \cite{nos_dijet}, are presented for comparison. As already demonstrated in Ref. \cite{nosdiphoton}, these two processes are subleading in comparison to the LbL one in the kinematical range considered. As expected for a resonance, the ALP production implies a peak in the invariant mass distribution. In addition, we have that for the production of an ALP with small mass ($m_a = 3.0$ GeV), the rapidity distributions for LHC and FCC energy are very similar to the LbL one. In contrast, for   $m_a = 15.0$ GeV, the distributions are strongly  suppressed and become similar to the Durham and DDP predictions, which implies that the inclusion of additional cuts is important to separate the ALP events.

In order to obtain realistic estimates for the ALP production in $PbPb$ collisions, which can be compared with the future experimental data, we will include in our analysis the experimental cuts  that are expected to be feasible in the next run of the LHC and in the future at the HE -- LHC and FCC. As in Ref. \cite{nosdiphoton}, we will consider two distinct configurations of cuts: one for a typical central detector, as ATLAS and CMS, and other for a forward detector, as  LHCb. The selection criteria implemented in our analysis are the following:   
\begin{itemize}
\item For a central detector: We will select events in which $m_{X}$ > 5 GeV and $E_{T}(\gamma,\gamma)$ > 2 GeV, where $E_T$ is the transverse energy of the photons. Moreover, we will impose a cut on  the acoplanarity ($1 -(\Delta \phi/\pi)$ < 0.01) and  transverse momentum of the diphoton system  ($p_{T}(\gamma,\gamma)$ < 0.1 GeV). Finally, we only will select events where photons are produced in the rapidity range $|\eta(\gamma^{1},\gamma^{2})| < 2.5$ with 0 extra tracks.

\item For a forward detector: We will select events in which $m_{X}$ > 1 GeV and $p_{T}(\gamma,\gamma)$ > 0.2 GeV, where $p_T$ is the transverse momentum of the photons. Moreover, we will impose a cut on  the acoplanarity ($1 -(\Delta \phi/\pi)$ < 0.01) and  transverse momentum of the diphoton system  ($p_{T}(\gamma,\gamma)$ < 0.1 GeV). Finally, we only will select events where photons are produced  in the rapidity range $2.0 < |\eta(\gamma^{1},\gamma^{2})| < 4.5$ with 0 extra tracks with $p_{T} > 0.1$ GeV in the rapidity range $−3.5 < \eta < −1.5$ and
$p_T > 0.5$ GeV in the range $−8.0 < \eta < −5.5$. Such set of cuts is considered in order to analyze the possibility of study the production of ALP's with  mass in the range $1 \le m_X \le 5$ GeV, which  cannot currently be reached by the central detectors. 
\end{itemize}

\begin{center}
\begin{table}
\begin{tabular}{|c|c|c|c|c|c|}
\hline 
{\bf $PbPb$ at $\sqrtsnn$ = 5.5 TeV} & {\bf LbL} & {\bf  Durham} & {\bf DDP} & {\bf $m_{a}=15$} GeV &{\bf $m_{a}=40$ GeV}  \tabularnewline
\hline 
Total Cross section {[}nb{]} &18000.0  & 167.0  & 17.7 & 11.0& 13.0  \tabularnewline
\hline 
$m_{X}> 5\:\rm{GeV}, E_{T}(\gamma,\gamma)>2\:\rm{GeV}$& 187.0  & 3.6   &17.7 & 11.0 &13.0   \tabularnewline
\hline 
$1- (\Delta \phi/\pi) < 0.01$ & 186.0  & 3.1 &6.9  &11.0 &13.0 \tabularnewline
\hline 
$p_{T}(\gamma\gamma)< 0.1$  GeV & 139.0 & 2.8 & 0.1 & 11.0 &13.0 \tabularnewline
\hline 
 $|\eta(\gamma,\gamma)|<2.5$ & 139 & 1.9 &  0.0&10.5 &12.5  \tabularnewline
\hline
\hline 
$13 < m\left(\gamma\gamma\right) < 17 $  & 7.3  & 0.1 &0.0  &8.6  &-  \tabularnewline
\hline 
$38 < m\left(\gamma\gamma\right) < 42 $  & 0.5 & 0.0 &0.0  &- & 11.5   \tabularnewline
\hline 
\hline 
{\bf $PbPb$ at $\sqrtsnn$ = 10.6 TeV} & {\bf LbL} & {\bf  Durham} & {\bf DDP} & {\bf $m_{a}=15$} GeV &{\bf $m_{a}=40$ GeV} \tabularnewline
\hline 
Total Cross section {[}nb{]} &27000.0  & 333.2  & 33.0 &21.7 & 35.1  \tabularnewline
\hline 
$m_{X}> 5\:\rm{GeV}, E_{T}(\gamma,\gamma)>2\:\rm{GeV}$& 352.9  & 7.6   &13.5 & 20.9 &  35.0  \tabularnewline
\hline 
$1- (\Delta \phi/\pi) < 0.01$ & 352.8  & 6.7 & 0.1 &20.9 & 35.0 \tabularnewline
\hline 
$p_{T}(\gamma\gamma)< 0.1$  GeV & 350.2 & 5.8 & 0.0 & 20.7 & 34.4 \tabularnewline
\hline 
 $|\eta(\gamma,\gamma)|<2.5$ & 227.6 &3.6  & 0.0 &15.1 & 28.8  \tabularnewline
\hline
\hline 
$13 < m\left(\gamma\gamma\right) < 17 $  & 20.0  &3.6  & 0.0 & 15.1 &-  \tabularnewline
\hline 
$38 < m\left(\gamma\gamma\right) < 42 $  &0.0  & 0.0 & 0.0 &- &   28.8 \tabularnewline
\hline 
\hline 
{\bf $PbPb$ at $\sqrtsnn$ = 39 TeV} & {\bf LbL} & {\bf  Durham} & {\bf DDP} & {\bf $m_{a}=15$} GeV &{\bf $m_{a}=40$ GeV} \tabularnewline
\hline 
Total Cross section {[}nb{]} & 52000.0 &  380 & 30.0 & 61.0 & 140.0 \tabularnewline
\hline 
$m_{X}> 5\:\rm{GeV}, E_{T}(\gamma,\gamma)>2\:\rm{GeV}$& 844.0  & 9.2  &13.0 & 58.8 & 140.0  \tabularnewline
\hline 
$1- (\Delta \phi/\pi) < 0.01$ & 840.0 & 8.0 &0.1  & 58.8& 139.0\tabularnewline
\hline 
$p_{T}(\gamma\gamma)< 0.1$  GeV & 836.0 & 7.0 & 0.0 & 58.0 & 139.0\tabularnewline
\hline 
 $|\eta(\gamma,\gamma)|<2.5$ & 431.0 & 3.4 &  0.0& 33.7& 93.0 \tabularnewline
\hline
\hline 
$13 < m\left(\gamma\gamma\right) < 17 $  &27.8  & 0.1 & 0.0 &33.7  & - \tabularnewline
\hline 
$38 < m\left(\gamma\gamma\right) < 42 $  & 1.5 & 0.0 & 0.0 &- &  93.0 \tabularnewline
\hline 
\end{tabular}
\caption{Predictions for the cross sections associated to the  ALP, LbL, Durham and double diffractive production (DDP) processes  after the inclusion of the exclusivity cuts for a typical central detector.}
\label{tab:central}
\end{table}
\end{center}

\begin{center}
\begin{table}
\begin{tabular}{|c|c|c|c|c|c|c|c|}
\hline 
{\bf $PbPb$ at $\sqrtsnn$ = 5.5 TeV} & {\bf LbL} & {\bf Durham} & {\bf DDP} & {\bf $m_{a}=3$ GeV} & {\bf$m_{a}=5$ GeV}  & {\bf $m_{a}=15$ GeV} & {\bf $m_{a}=40$ GeV}\tabularnewline
\hline 
Total Cross section {[}nb{]} & 18000.0 & 167.0  & 17.7 & 13000.0& 363.0 & 11.0 & 13.0\tabularnewline
\hline 
$m_{X}> 1\:\rm{GeV}, p_{T}(\gamma,\gamma)>0.2\:\rm{GeV}$& 13559.0 & 142.0 &17.6 &12873.0 & 360.0 & 11.0 & 13.0 \tabularnewline
\hline 
$1- (\Delta \phi/\pi) < 0.01$ & 8834.0 & 51.0 & 0.2 &11033.0 &335.0 & 11.0 &13.0\tabularnewline
\hline 
$p_{T}(\gamma\gamma)< 0.1$  GeV & 8826.0 & 47.0 &0.0 & 11019.0 &334.7 &10.8 &13.0\tabularnewline
\hline 
 $2.0<\eta(\gamma,\gamma)<4.5$ &  616.0&  3.7& 0.0 &974.0 & 23.4 &0.2 &0.02\tabularnewline
\hline
\hline
$2 < m\left(\gamma\gamma\right) < 4 $  &83.7  &3.2  & 0.0 &974.0 &- & -& -\tabularnewline
\hline
$5 < m\left(\gamma\gamma\right) < 7 $  &32.0  &1.0  & 0.0& -& 23.4 & -&-\tabularnewline
\hline 
$13 < m\left(\gamma\gamma\right) < 17 $  &0.0  & 0.0 &  0.0& - &- & 0.2&- \tabularnewline
\hline 
$38 < m\left(\gamma\gamma\right) < 42 $  & 0.0 & 0.0 & 0.0 & -& - & - & 0.02 \tabularnewline
\hline 
\hline 
{\bf $PbPb$ at $\sqrtsnn$ = 10.6 TeV} & {\bf LbL} & {\bf Durham} & {\bf DDP} & {\bf $m_{a}=3$ GeV} & {\bf$m_{a}=5$ GeV}  & {\bf $m_{a}=15$ GeV} & {\bf $m_{a}=40$ GeV}\tabularnewline
\hline 
Total Cross section {[}nb{]} & 27000.0 & 333.2  &33.0 &21000.0 &587.4 &21.7  &35.1 \tabularnewline
\hline 
$m_{X}> 1\:\rm{GeV}, p_{T}(\gamma,\gamma)>0.2\:\rm{GeV}$&20372.9  & 284.6  &33.0 &20793.3 & 585.2  & 21.7 & 35.1 \tabularnewline
\hline 
$1- (\Delta \phi/\pi) < 0.01$ &13958.5  & 103.2 & 0.3 & 18190.3 &554.8 &21.6  &35.1\tabularnewline
\hline 
$p_{T}(\gamma\gamma)< 0.1$  GeV & 13949.0 & 95.1 & 0.0 & 18171.6  &553.7 &21.4 &34.6\tabularnewline
\hline 
 $2.0<\eta(\gamma,\gamma)<4.5$ & 1069.5 & 8.3 & 0.0 &1904.6 & 52.1 &1.0 &0.4\tabularnewline
\hline
\hline
$2 < m\left(\gamma\gamma\right) < 4 $  &159.3  & 7.1 & 0.0 &1904.6 &- & -& -\tabularnewline
\hline
$5 < m\left(\gamma\gamma\right) < 7 $  &69.1  & 2.3 & 0.0& -& 52.1  & -&-\tabularnewline
\hline 
$13 < m\left(\gamma\gamma\right) < 17 $  & 0.8 & 0.0 &0.0  & - &- &1.0 &- \tabularnewline
\hline 
$38 < m\left(\gamma\gamma\right) < 42 $  &0.0  & 0.0 & 0.0 & -& - & - & 0.4 \tabularnewline
\hline 
\hline 
{\bf $PbPb$ at $\sqrtsnn$ = 39 TeV} & {\bf LbL} & {\bf Durham} & {\bf DDP} & {\bf $m_{a}=3$ GeV} & {\bf$m_{a}=5$ GeV}  & {\bf $m_{a}=15$ GeV} & {\bf $m_{a}=40$ GeV}\tabularnewline
\hline 
Total Cross section {[}nb{]} & 52000.0 & 380.0   & 30.0  & 43000.0 & 1300.0  & 61.0  &140.0 \tabularnewline
\hline 
$m_{X}> 1\:\rm{GeV}, p_{T}(\gamma,\gamma)>0.2\:\rm{GeV}$& 38025.0 & 325.0&30.0 &42587.0 &  1295.0& 61.0 & 140.0 \tabularnewline
\hline 
$1- (\Delta \phi/\pi) < 0.01$ &28216.0  &118.0  & 0.3 &38320.0 &1243.0 &61.0  &140.0\tabularnewline
\hline 
$p_{T}(\gamma\gamma)< 0.1$  GeV & 28202.0 & 109.0 &0.0 & 38290.0 &1241.0 &60.0 &139.0\tabularnewline
\hline 
 $2.0<\eta(\gamma,\gamma)<4.5$ & 2229.0 &10.0  & 0.0 &4377.0 & 139.0 &5.8 &8.7\tabularnewline
\hline
\hline
$2 < m\left(\gamma\gamma\right) < 4 $  &  383.0 &7.7  &0.0  &4377.0 &- & -& -\tabularnewline
\hline
$5 < m\left(\gamma\gamma\right) < 7 $  & 176.0 &3.0 &0.0 & -& 139.0 & -&-\tabularnewline
\hline 
$13 < m\left(\gamma\gamma\right) < 17 $ &  4.5& 0.0 & 0.0 & - &- &5.8 &- \tabularnewline
\hline 
$38 < m\left(\gamma\gamma\right) < 42 $  & 0.2 & 0.0 &0.0  & -& - & - &  8.7\tabularnewline
\hline 
\end{tabular}
\caption{Predictions for the cross sections associated to the  ALP, LbL, Durham and double diffractive production (DDP) processes  after the inclusion of the exclusivity cuts for a typical forward detector.}
\label{tab:forward}
\end{table}
\end{center}

Our predictions for the central and forward configurations are presented in Tables \ref{tab:central} and \ref{tab:forward}, respectively. The inclusion of the exclusivity cuts strongly reduces the background, with the DDP contribution being fully eliminated and the Durham being of the order of 2 \%. For a central detector, we have that the selection in the invariant mass range around the ALP mass implies that the LbL background becomes of the order or smaller the ALP signal. In particular, for $m_a = 40$ GeV, the LbL background becomes negligible and the study of the diphoton production is a direct probe of the ALP.  On the other hand, the results presented in Table \ref{tab:forward} for a forward detector indicate that it is ideal to probe an ALP with small mass. {  Considering the expected luminosities for the next run of the LHC and future colliders, which are 10 nb$^{-1}$ and 110 nb$^{-1}$, the associated number of ALP events are, respectively,  $\approx$  19046 and 481470, for $m_a = 3.0$ GeV. Also, taking into account the integrated luminosity achieved by the LHCb in 2018, $210\mu b^{-1}$, the significance values obtained for $\sqrt{s}=5.5, \,10.6$ and $39$ TeV are $\approx 48, \,68$ and $102$, respectively. The expected luminosities for $\sqrt{s}=$10.6 and 39 TeV relative to $5\sigma$ are $0.1\:\mu b^{-1}$ and $0.004\:\mu b^{-1}$.} Such results demonstrate the potentiality of the LHCb detector to constrain the main properties of the ALP.

\begin{center}
 \begin{figure}[t]
 \includegraphics[width=0.45\textwidth]{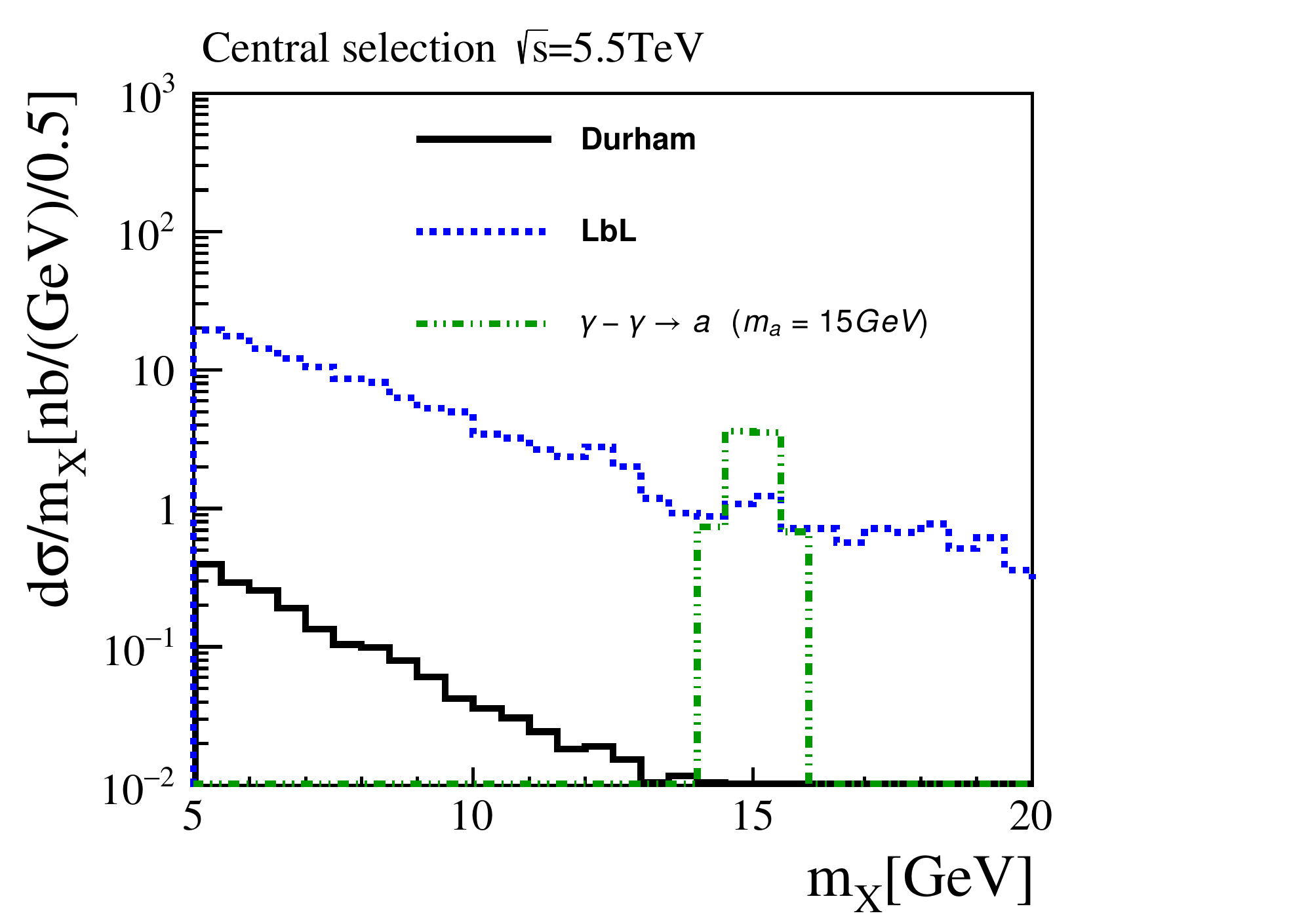}
  \includegraphics[width=0.45\textwidth]{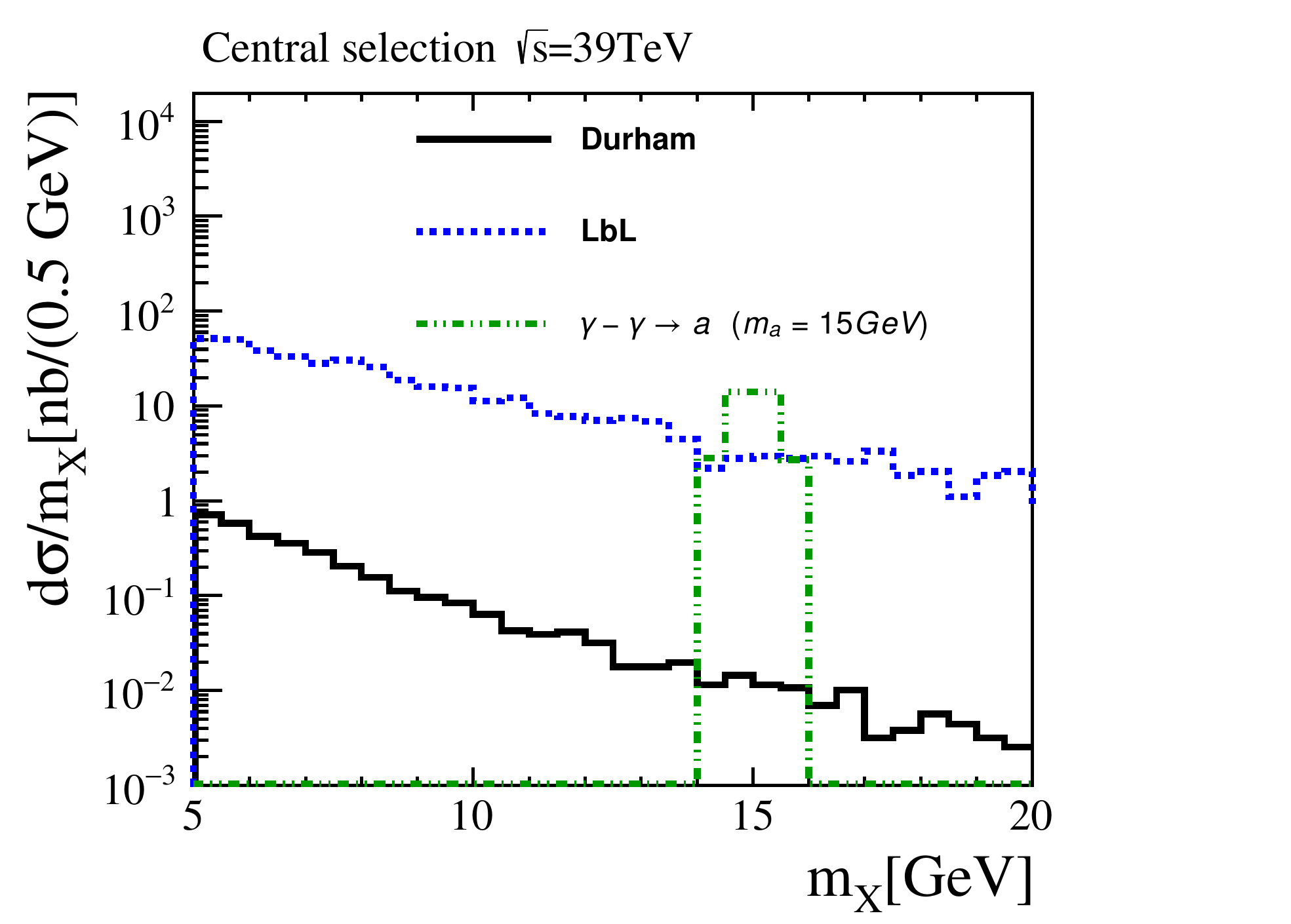}
 \includegraphics[width=0.45\textwidth]{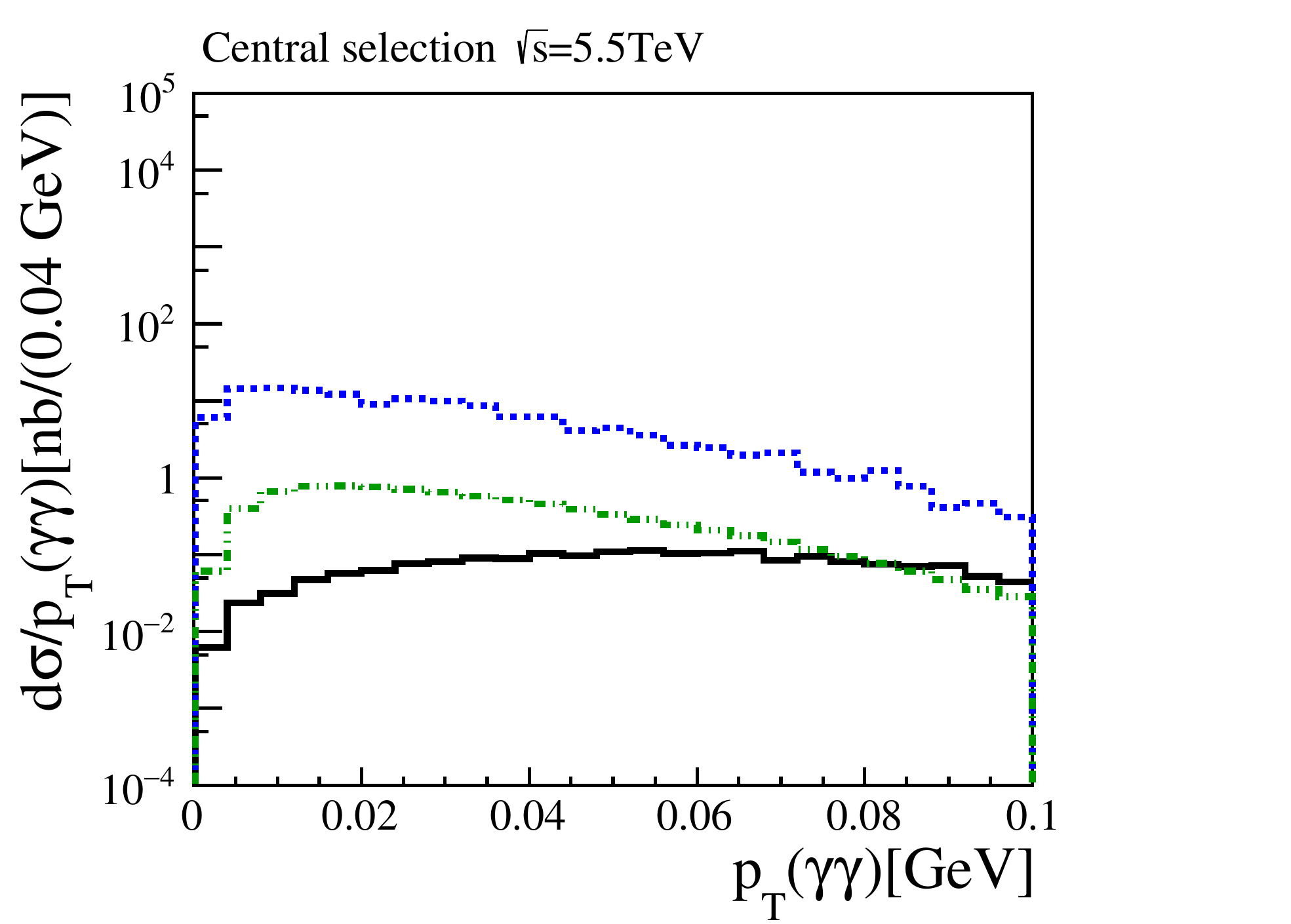}
  \includegraphics[width=0.45\textwidth]{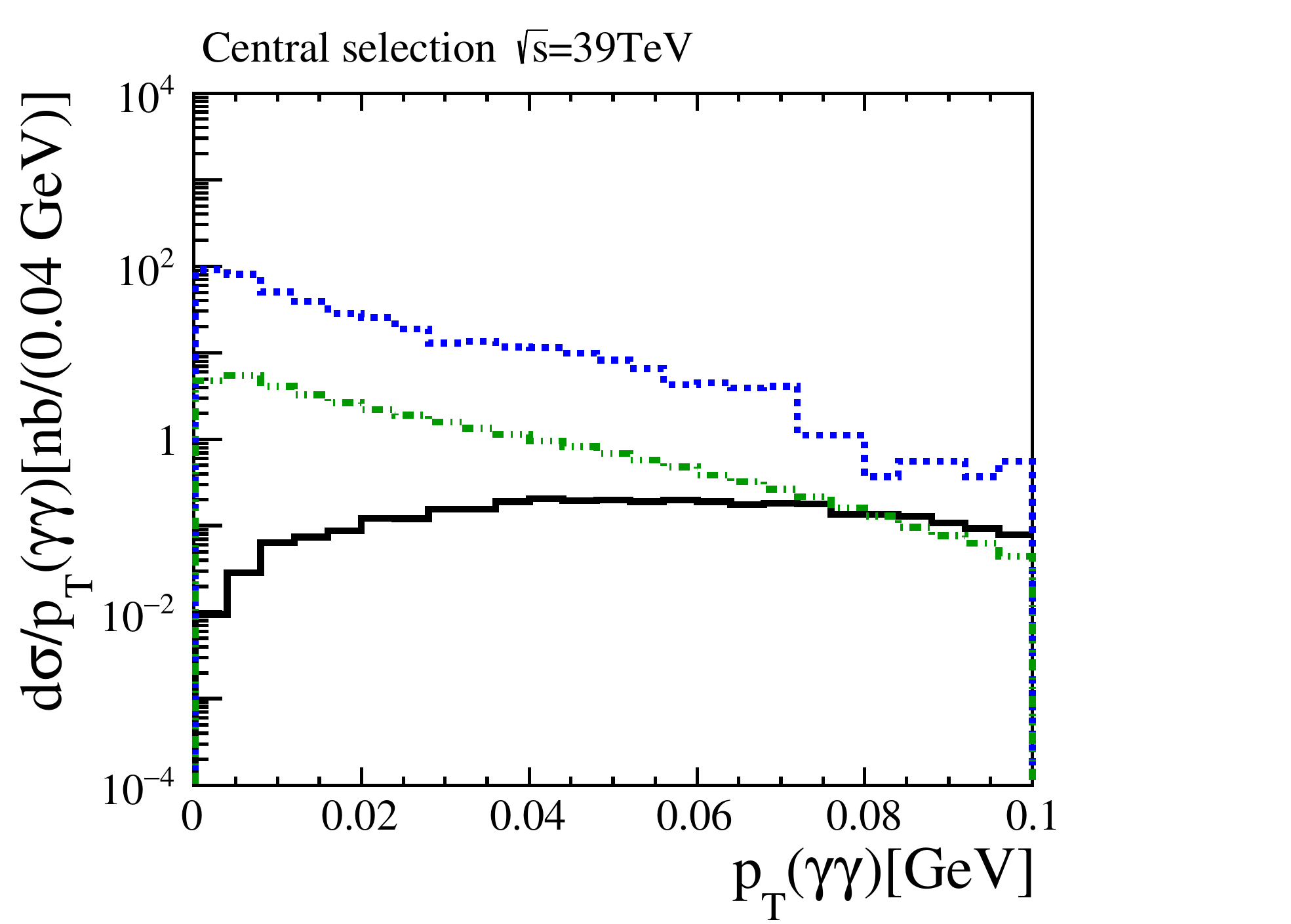}
 \includegraphics[width=0.45\textwidth]{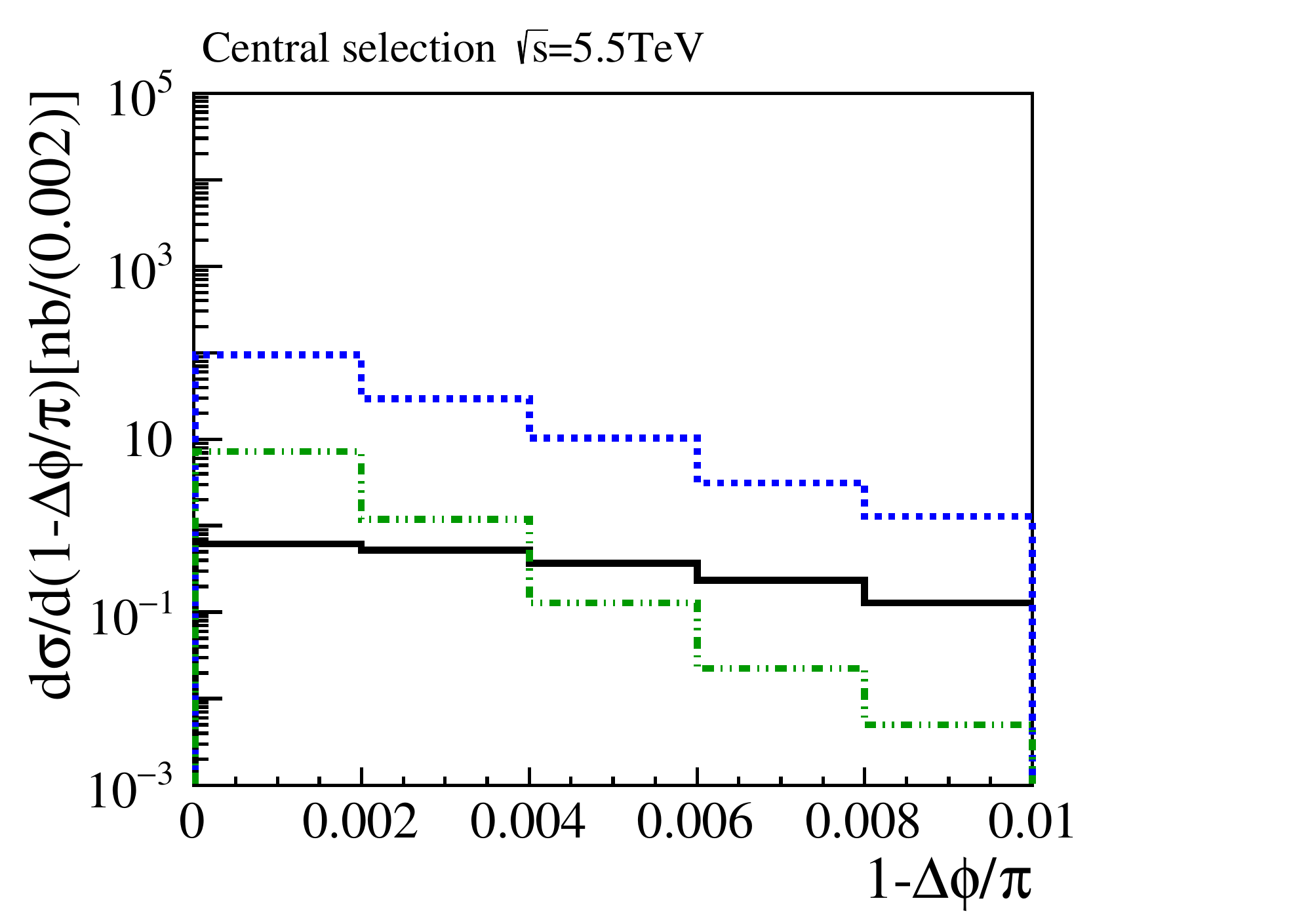}
 \includegraphics[width=0.45\textwidth]{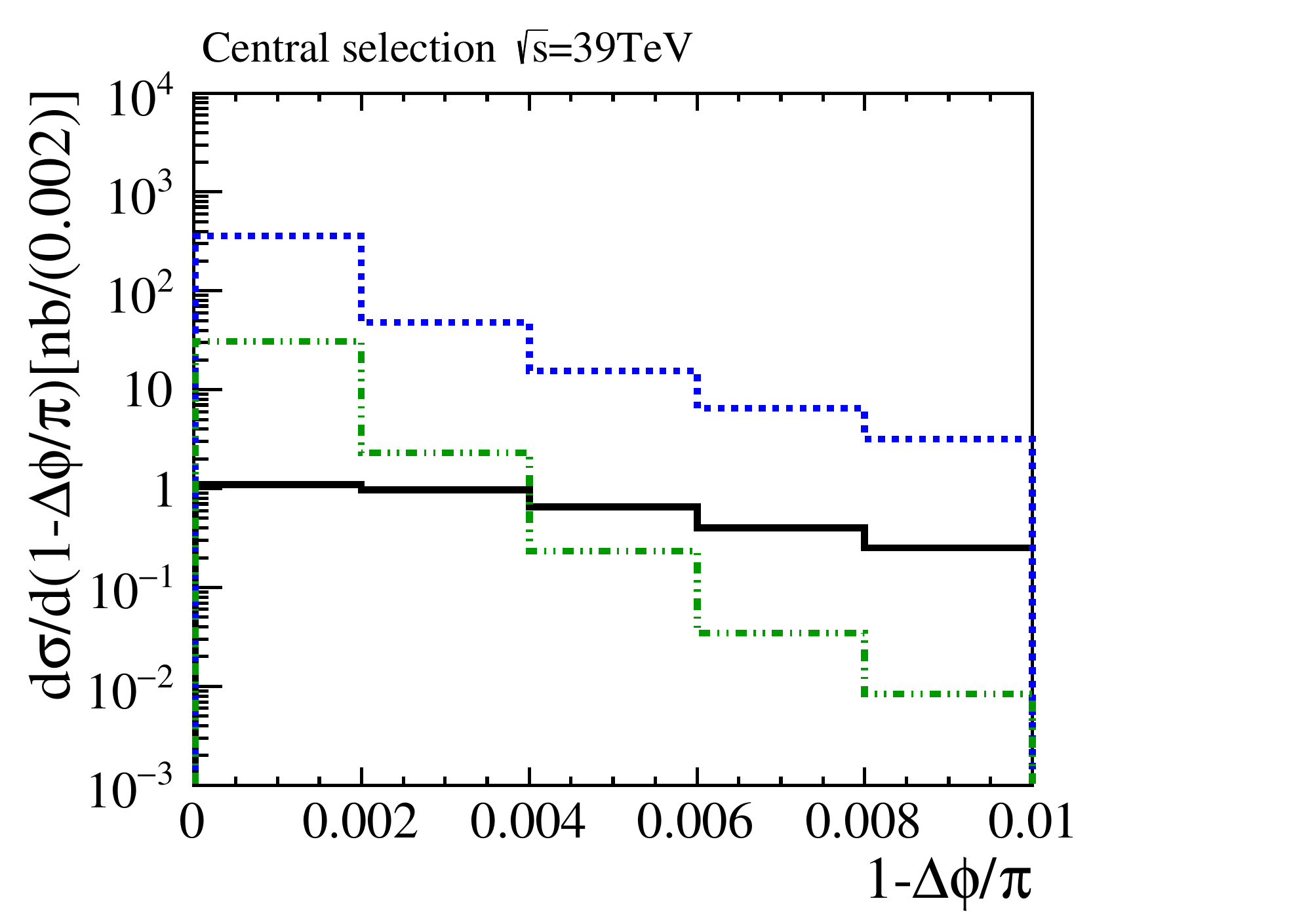}
 \includegraphics[width=0.45\textwidth]{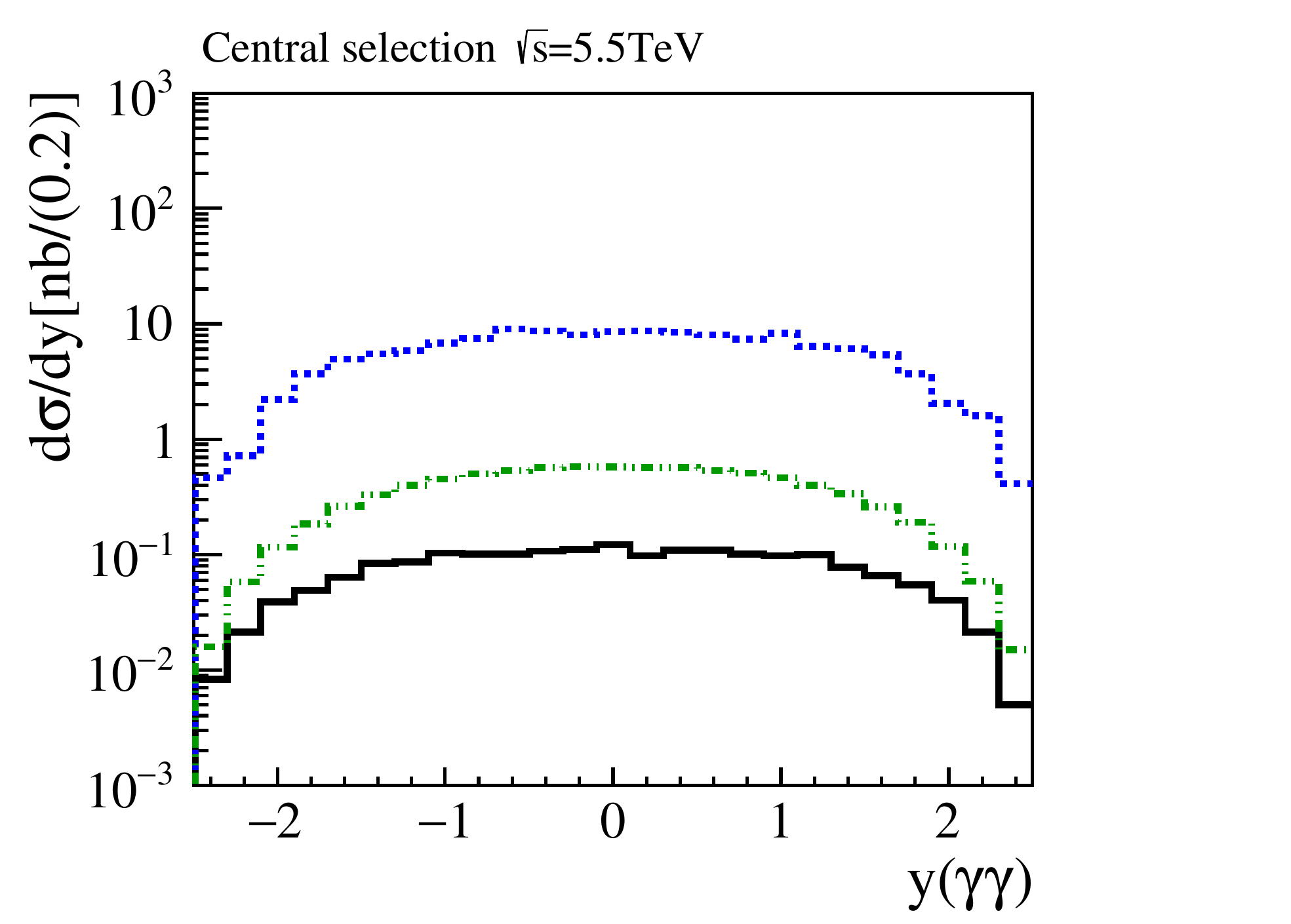}
  \includegraphics[width=0.45\textwidth]{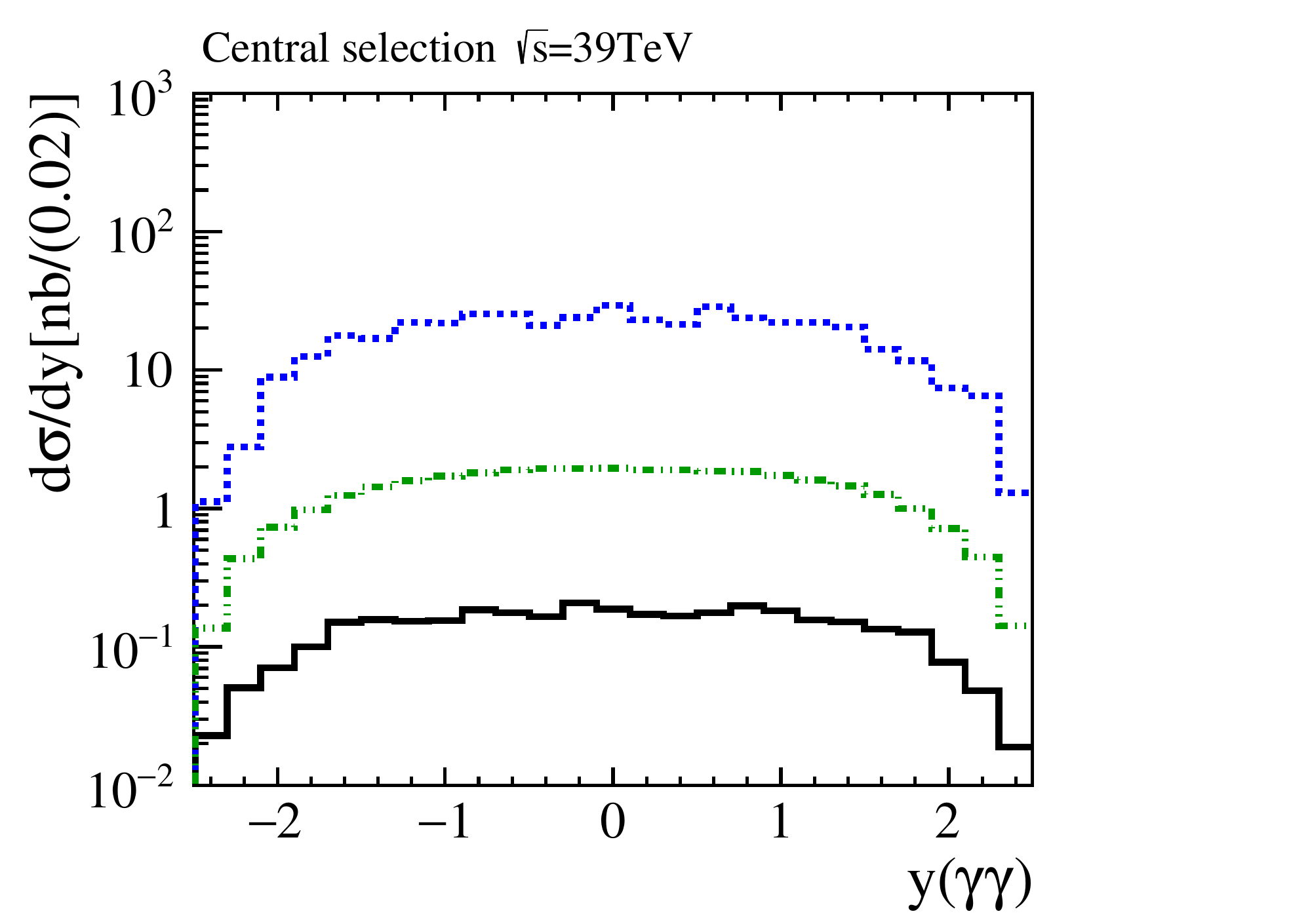}
  \caption{Differential cross sections  of the diphoton system for invariant mass $m_{X}$, transverse momentum $p_{T}(\gamma\gamma)$, rapidity $y({\gamma\gamma})$ and acoplanarity for LHC (left panels) and FCC (right panels) energies considering a central detector without the inclusion of a cut on the ALP mass.}
\label{fig:central}
 \end{figure}
 \end{center}

\begin{center}
 \begin{figure}[t]
 \includegraphics[width=0.45\textwidth]{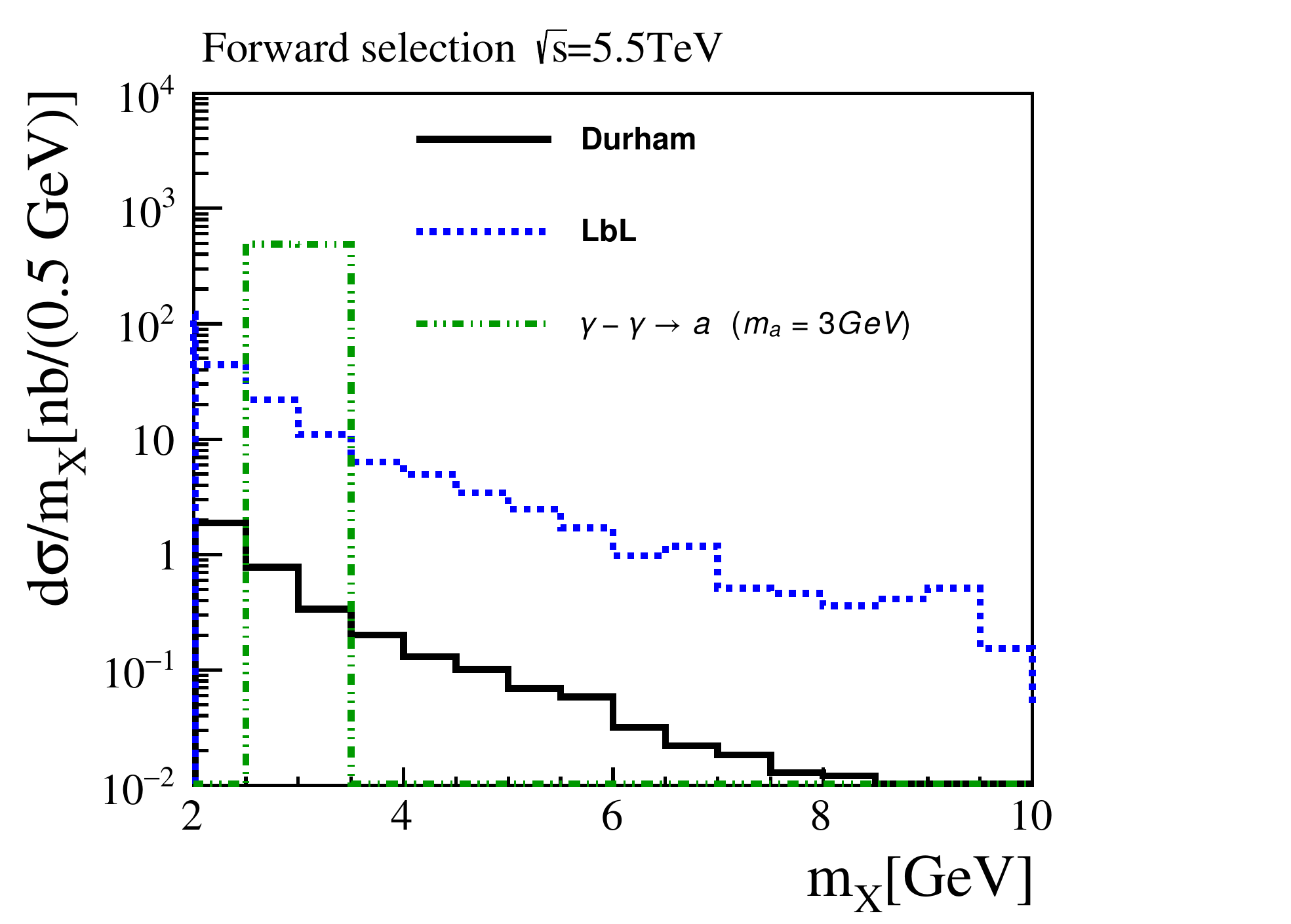}
  \includegraphics[width=0.45\textwidth]{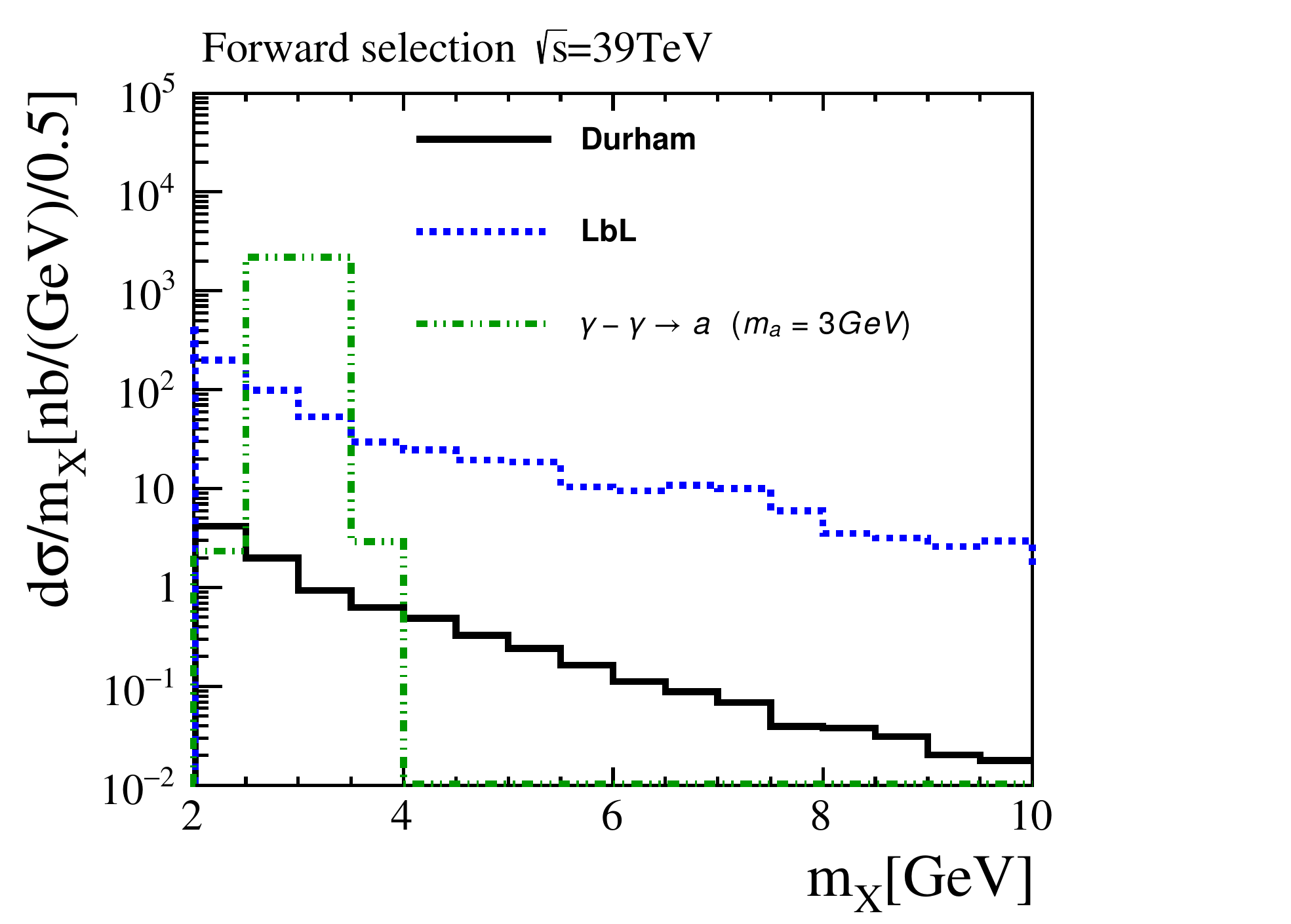}
 \includegraphics[width=0.45\textwidth]{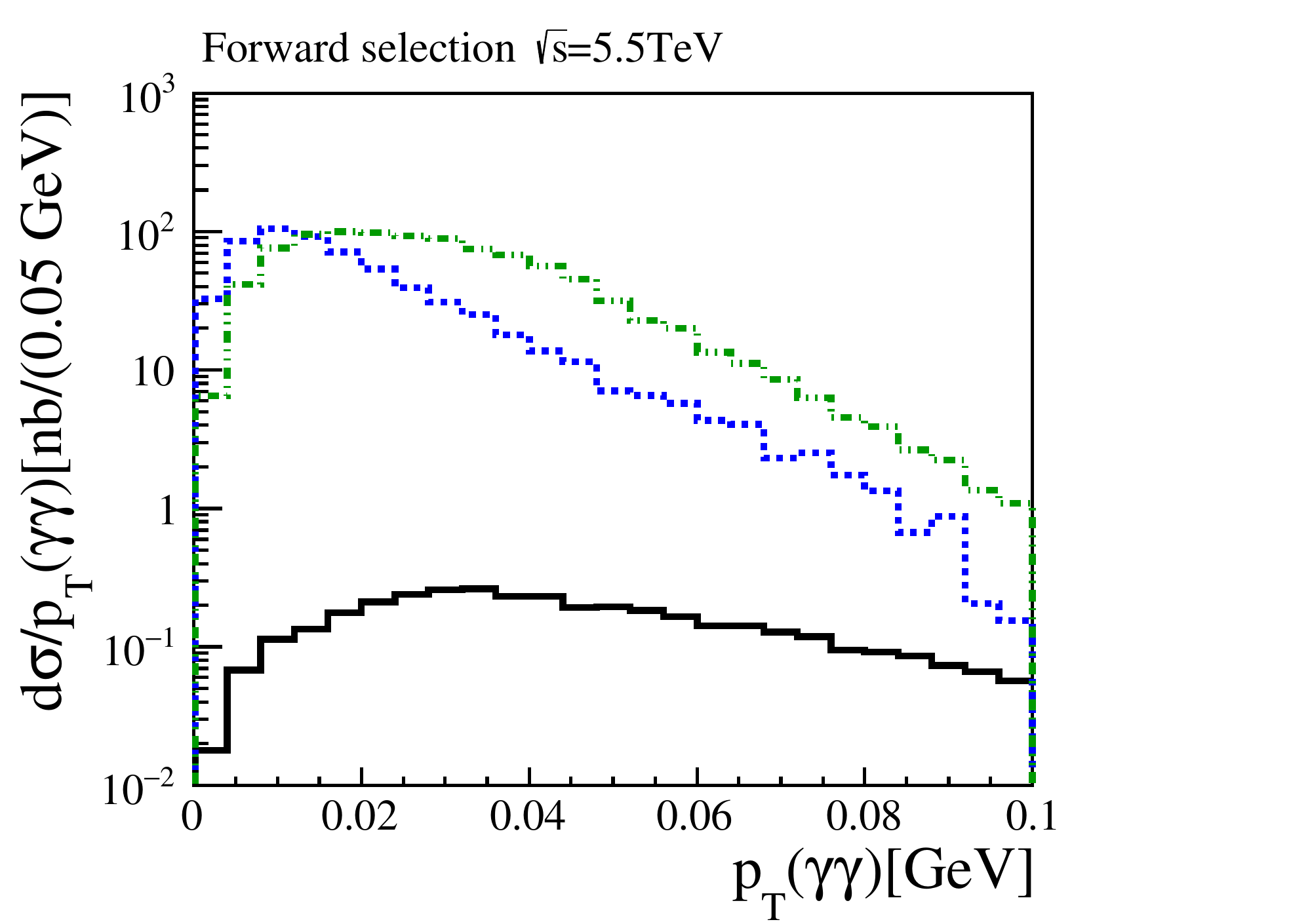}
 \includegraphics[width=0.45\textwidth]{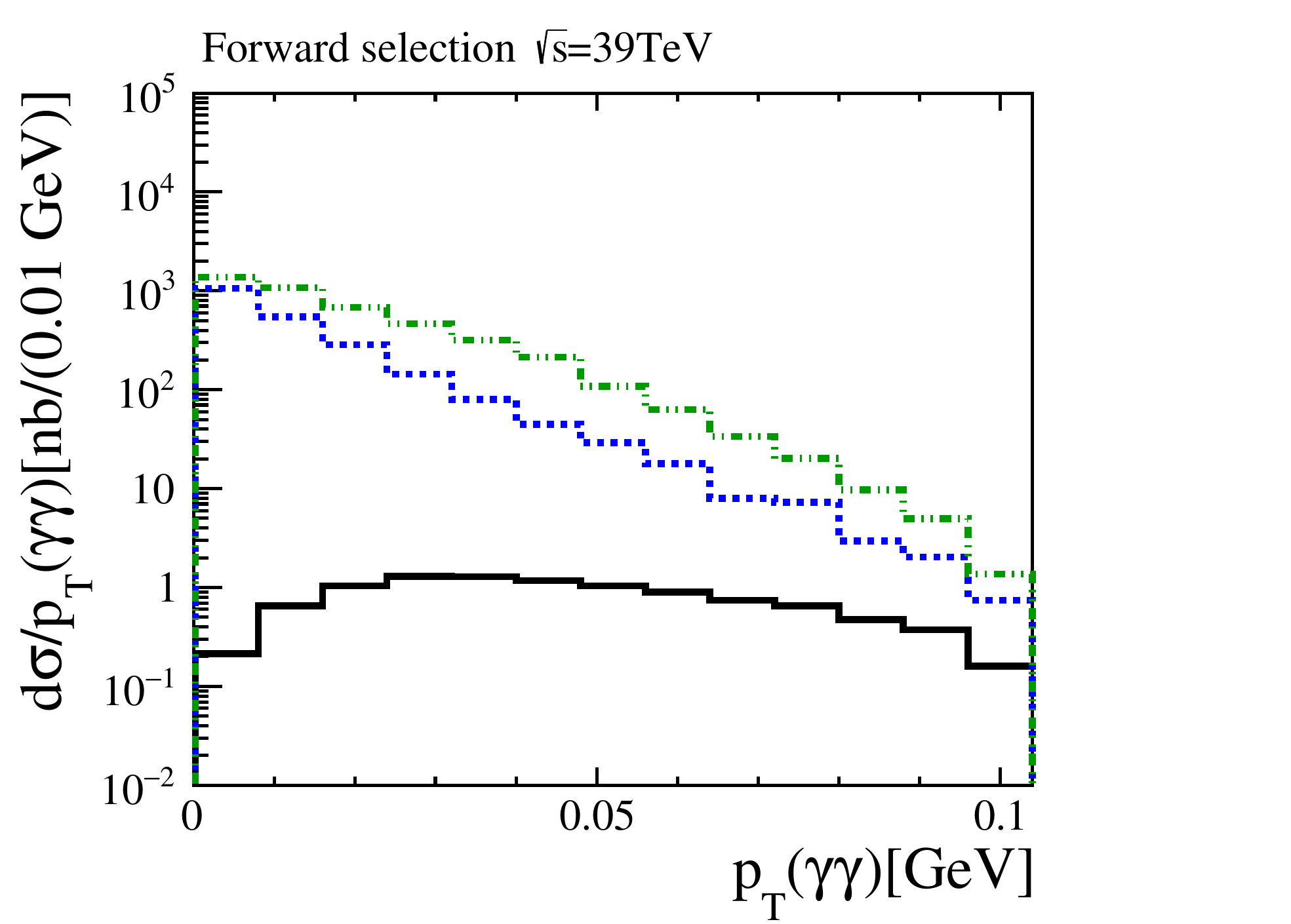}
 \includegraphics[width=0.45\textwidth]{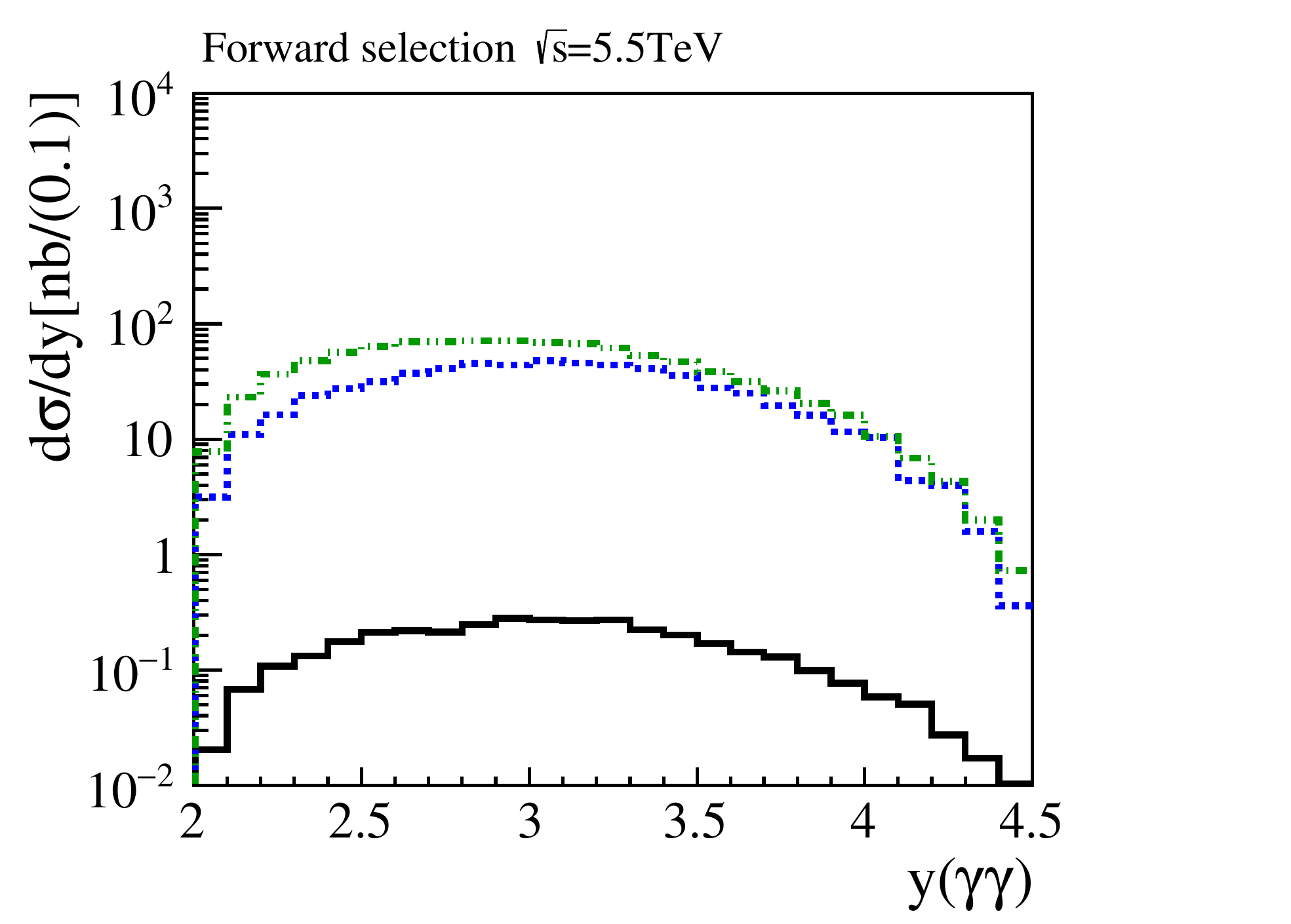}
  \includegraphics[width=0.45\textwidth]{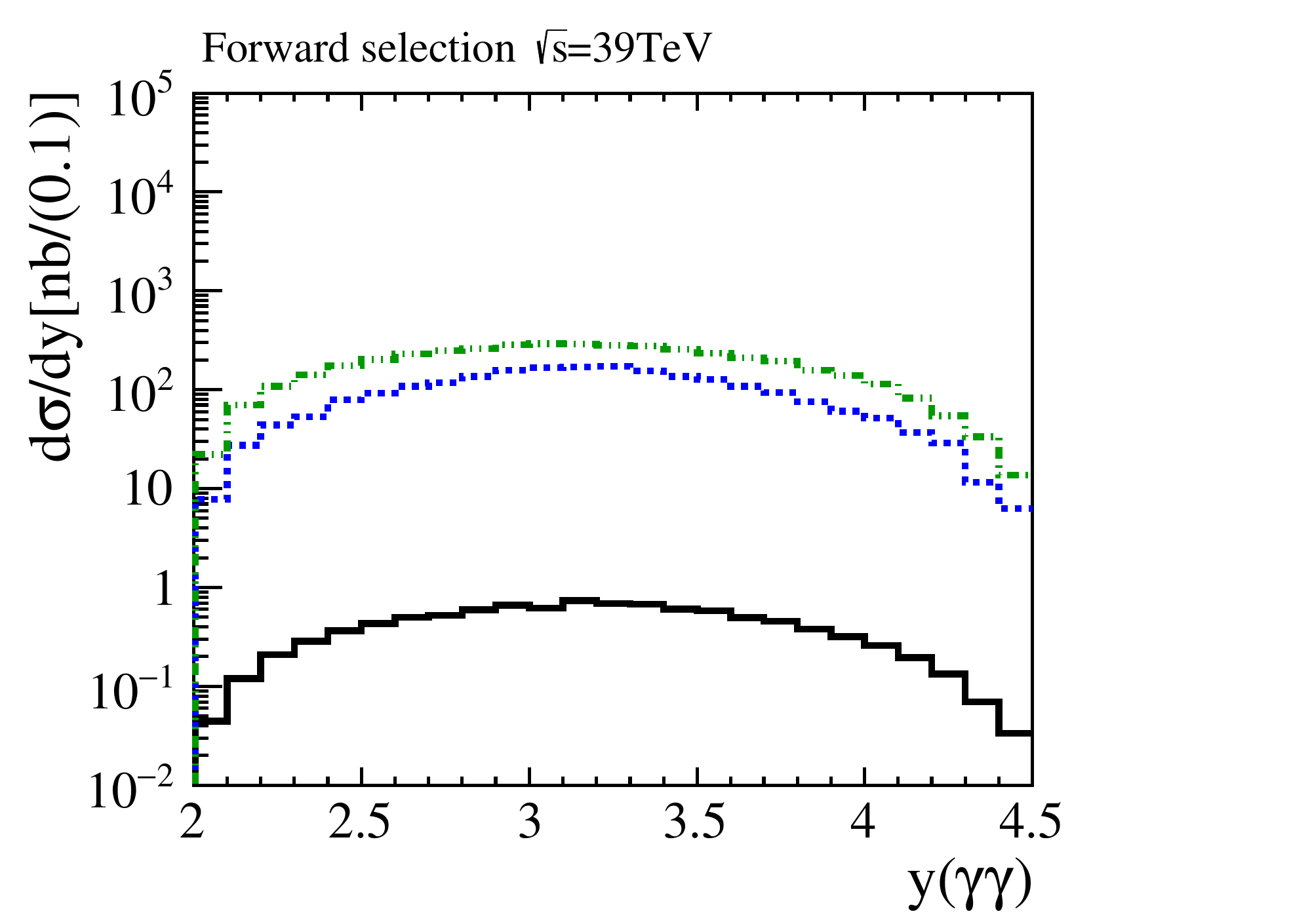}
 \includegraphics[width=0.45\textwidth]{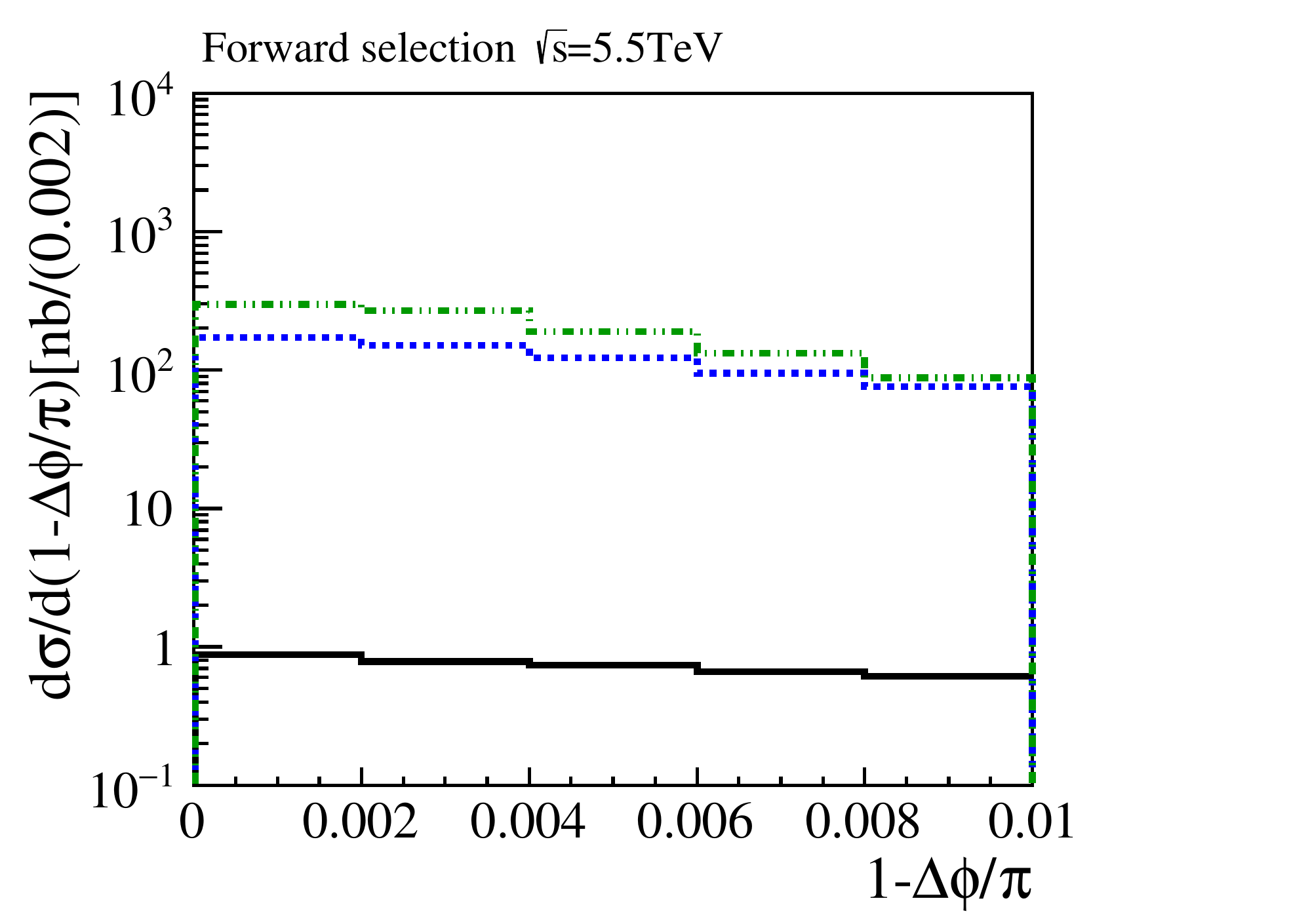}
 \includegraphics[width=0.45\textwidth]{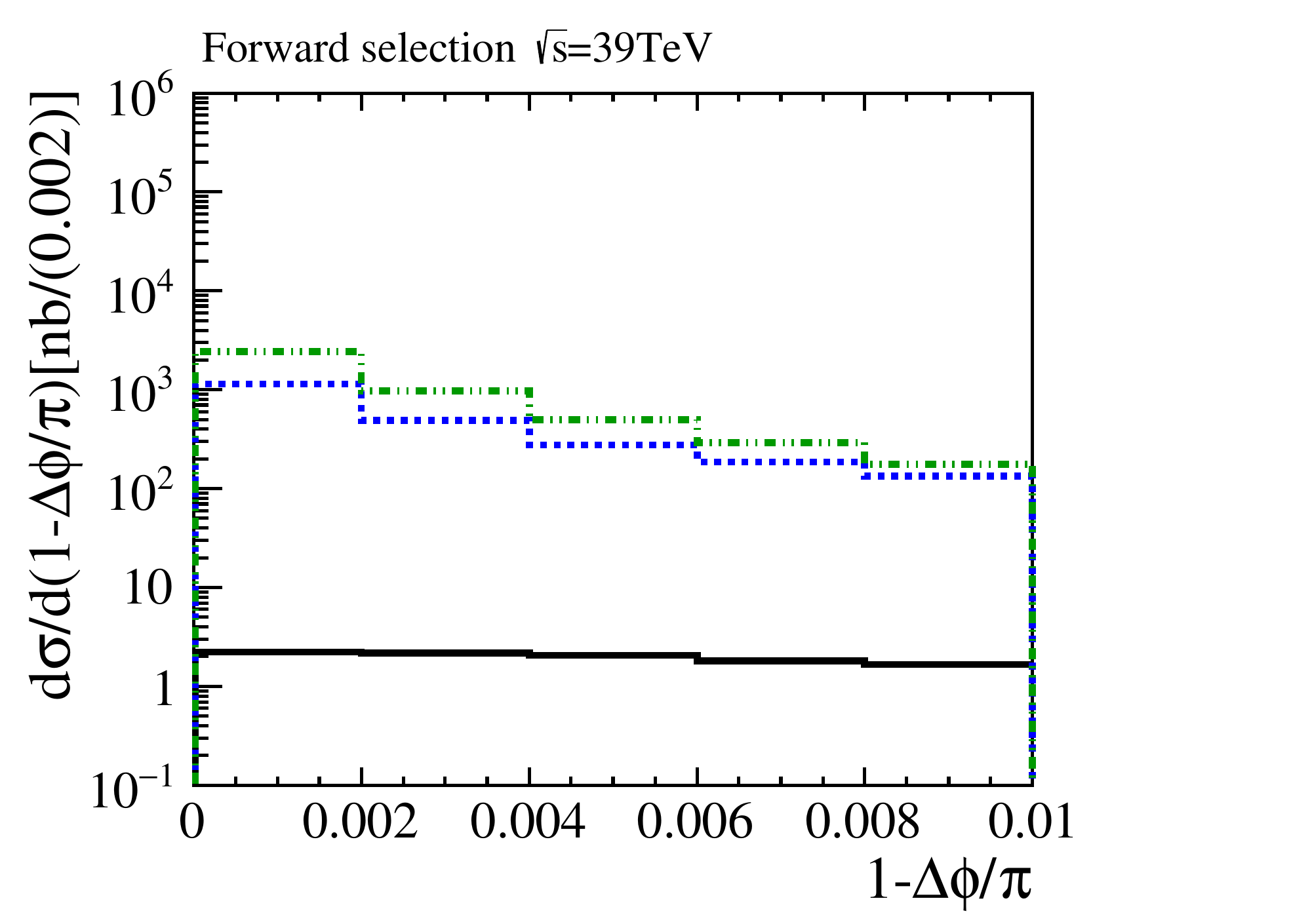}
 \caption{Differential cross sections  of the diphoton system for invariant mass $m_{X}$, transverse momentum $p_{T}(\gamma\gamma)$, rapidity $y({\gamma\gamma})$ and acoplanarity for LHC (left panels) and FCC (right panels) energies considering a forward detector without the inclusion of a cut on the ALP mass.}
\label{fig:forward}
 \end{figure}
 \end{center}

In Fig. \ref{fig:central}  we present our predictions for the  invariant mass $m_X$, transverse momentum $p_T (\gamma \gamma)$, rapidity $y(\gamma \gamma)$ and acoplanarity distributions considering a central detector, $m_a = 15$ GeV, the exclusivity cuts discussed before and  $PbPb$ collisions at the LHC (left panels) and FCC (right panels). These results have been derived before the selection in the invariant mass of the diphoton system. We have that the contribution of the  Durham process is, in general, negligible, only becoming competitive for a diphoton with a large transverse momentum. The predictions for the LbL background are approximately one order of magnitude larger than ALP signal, but the shape of the $p_T (\gamma \gamma)$, $y(\gamma \gamma)$ and acoplanarity distributions are similar.  On the other hand, for a forward detector and assuming that $m_a = 3.0$ GeV, the results presented in  Fig. \ref{fig:forward} indicate that the LbL and ALP predictions for the distributions are very similar, with the ALP one being slightly larger.

  \section{Summary}
 \label{sec:conc}
The high photon -- photon luminosity present in ultraperipheral heavy -- ion collisions become feasible  the search of New Physics in photon -- induced interactions. 
 One of more interesting final states is the diphoton system with a small invariant mass, which is dominantly produced by the  Light -- by -- Light scattering and can also be generated  by an ALP resonance  in the $s$ -- channel. In this paper we have performed an exploratory study of the ALP production in $PbPb$ collisions at the LHC, HE -- LHC and FCC energies, considering four combinations for the ALP mass and coupling and taking into account the acceptance of the LHC detectors. In particular, a detailed analysis of the ALP production in the kinematical range probed by the LHCb detector was performed by the first time.  Our results demonstrated that the LbL background can be strongly reduced  by the exclusivity cuts and that the ALP signal is dominant for a forward detector. Consequently, a future experimental anaysis of the diphoton final state is a promissing observable to probe the existence of the Axionlike particles  and its properties.

\begin{acknowledgments}
VPG acknowledge very useful discussions about photon - induced interactions with Gustavo Gil da Silveira, Mariola Klusek-Gawenda and Antoni Szczurek.
This work was  partially financed by the Brazilian funding
agencies CNPq, CAPES,  FAPERGS, FAPERJ and INCT-FNA (processes number 
464898/2014-5 and 88887.461636/2019-00).
\end{acknowledgments}


\begin{thebibliography}{99}

\bibitem{Jaeckel:2015jla} 
  J.~Jaeckel and M.~Spannowsky,
  Phys.\ Lett.\ B {\bf 753}, 482 (2016).

\bibitem{Bauer:2017ris} 
  M.~Bauer, M.~Neubert and A.~Thamm,
  JHEP {\bf 1712}, 044 (2017).


\bibitem{knapen} 
  S.~Knapen, T.~Lin, H.~K.~Lou and T.~Melia,
  Phys.\ Rev.\ Lett.\  {\bf 118}, no. 17, 171801 (2017).


\bibitem{Aloni:2018vki} 
  D.~Aloni, Y.~Soreq and M.~Williams,
  Phys.\ Rev.\ Lett.\  {\bf 123}, no. 3, 031803 (2019).

\bibitem{royon} 
  C.~Baldenegro, S.~Hassani, C.~Royon and L.~Schoeffel,
  Phys.\ Lett.\ B {\bf 795}, 339 (2019).



\bibitem{Aloni:2019ruo} 
  D.~Aloni, C.~Fanelli, Y.~Soreq and M.~Williams,
  Phys.\ Rev.\ Lett.\  {\bf 123}, no. 7, 071801 (2019).
  
\bibitem{Bauer:2018uxu} 
  M.~Bauer, M.~Heiles, M.~Neubert and A.~Thamm,
  Eur.\ Phys.\ J.\ C {\bf 79}, no. 1, 74 (2019).  
  
  \bibitem{Yue:2019gbh} 
  C.~X.~Yue, M.~Z.~Liu and Y.~C.~Guo,
  Phys.\ Rev.\ D {\bf 100}, no. 1, 015020 (2019).
  
  \bibitem{Ebadi:2019gij} 
  J.~Ebadi, S.~Khatibi and M.~Mohammadi Najafabadi,
  Phys.\ Rev.\ D {\bf 100}, no. 1, 015016 (2019).
  
  
  
  \bibitem{Alves:2019xpc} 
  A.~Alves, A.~G.~Dias and D.~D.~Lopes,
  arXiv:1911.12394 [hep-ph].
  
  
\bibitem{upc1}
C. A. Bertulani and G. Baur, { Phys. Rep.} {\bf 163}, 299 (1988).

\bibitem{upc2}
F.~Krauss, M.~Greiner and G.~Soff,
  Prog.\ Part.\ Nucl.\ Phys.\  {\bf 39}, 503 (1997).

\bibitem{upc3}
    G.~Baur, K.~Hencken and D.~Trautmann,
  J.\ Phys.\ G {\bf 24}, 1657 (1998).
  
  \bibitem{upc4}
 G. Baur, K. Hencken, D. Trautmann, S. Sadovsky, Y. Kharlov, Phys.
Rep. {\bf 364}, 359 (2002).

\bibitem{upc5}
   C.~A. Bertulani, S.~R.~Klein and J.~Nystrand, Ann. Rev. Nucl. Part. Sci. {\bf 55}, 
271 (2005).

\bibitem{upc6}
 V.~P.~Goncalves and M.~V.~T.~Machado,
  J.\ Phys.\ G {\bf 32}, 295 (2006).

\bibitem{upc7}
      A.~J.~Baltz {\it et al.},
  Phys.\ Rept.\  {\bf 458}, 1 (2008).

\bibitem{upc8}
      J.~G.~Contreras and J.~D.~Tapia Takaki,
  Int.\ J.\ Mod.\ Phys.\ A {\bf 30}, 1542012 (2015).

\bibitem{upc9}
      K.~Akiba {\it et al.} [LHC Forward Physics Working Group],
  J.\ Phys.\ G {\bf 43}, 110201 (2016).


\bibitem{epa} 
  V.~M.~Budnev, I.~F.~Ginzburg, G.~V.~Meledin and V.~G.~Serbo,
  Phys.\ Rept.\  {\bf 15}, 181 (1975).
  
  
\bibitem{nosdiphoton}
 R.~O.~Coelho, V.~P.~Goncalves, D.~E.~Martins and M.~S.~Rangel,
  arXiv:2002.03902 [hep-ph].

\bibitem{he_lhc} 
  A.~Abada {\it et al.} [FCC Collaboration],
  Eur.\ Phys.\ J.\ ST {\bf 228}, no. 5, 1109 (2019).


\bibitem{fcc} 
  A.~Abada {\it et al.} [FCC Collaboration],
  Eur.\ Phys.\ J.\ ST {\bf 228}, no. 4, 755 (2019).



\bibitem{superchic3} 
  L.~A.~Harland-Lang, V.~A.~Khoze and M.~G.~Ryskin,
  Eur.\ Phys.\ J.\ C {\bf 79}, no. 1, 39 (2019). 


\bibitem{nos_dijet}
E.~Basso,~V.~P.~Goncalves, A.~K.~Kohara, M.~S.~Rangel, Eur. Phys. J. C  {\bf 77}, 600 (2017).


\end{thebibliography}
 \end{document}